\documentclass[letter,nofootinbib,12pt,superscriptaddress]{revtex4-2}

\usepackage[colorlinks,citecolor=blue]{hyperref}
\usepackage{amsmath}
\usepackage{mathrsfs}
\usepackage{bbm}
\usepackage{amsfonts}
\usepackage{amssymb}
\usepackage{latexsym}
\usepackage{graphicx}
\usepackage[english]{babel}
\usepackage{multirow}
\usepackage{float}
\usepackage{url}
\usepackage{slashed}
\usepackage[compat=1.1.0]{tikz-feynman}
\usepackage{orcidlink}

\begin{document}

\title{Effects of stimulated emission and superradiant growth of non-spherical axion cluster}
 
\author{Liang Chen
\orcidlink{0000-0002-0224-7598}}
\email{bqipd@protonmail.com}
\affiliation{School of Fundamental Physics and Mathematical Sciences,
Hangzhou Institute for Advanced Study, UCAS, Hangzhou 310024, China}
\affiliation{University of Chinese Academy of Sciences, 100190 Beijing, China}

\author{Da Huang
\orcidlink{0000-0001-8357-754X}}
\email{dahuang@bao.ac.cn}
\affiliation{National Astronomical Observatories, Chinese Academy of Sciences, Beijing, 100012, China}
\affiliation{School of Fundamental Physics and Mathematical Sciences,
Hangzhou Institute for Advanced Study, UCAS, Hangzhou 310024, China}
\affiliation{University of Chinese Academy of Sciences, 100190 Beijing, China}
\affiliation{International Centre for Theoretical Physics Asia-Pacific, Beijing/Hangzhou, China}

\author{Chao-Qiang Geng
\orcidlink{0000-0002-8895-9245} }
\email{cqgeng@ucas.ac.cn}
\affiliation{School of Fundamental Physics and Mathematical Sciences,
Hangzhou Institute for Advanced Study, UCAS, Hangzhou 310024, China}
\affiliation{International Centre for Theoretical Physics Asia-Pacific, Beijing/Hangzhou, China}

\date{\today}

\begin{abstract}
\noindent We explore the stimulated emission of photons in non-spherical axion clusters with or without the axion source from the superradiance of a rotating black hole (BH). In particular, we focus on the cluster with the initial axion distribution in the $(l,m)=(1,1)$ mode  which mimics the shape of an axion cloud induced by the BH superradiance. After establishing the hierarchy of Boltzmann equations governing a general non-spherical axion-photon system, we examine the evolution of photon and axion distributions in the cluster and possible stimulated emission signals. In the case without the axion source, the resultant signal would be a single photon pulse. As for the system with the BH superradiance as the axion source, multiple pulses are predicted. We also show that, for the latter case, the combined effects of stimulated emissions and the axion production from the BH superradiance could reach a balance where the axion cluster becomes uniformly and spherically distributed. 
Due to the energy and temporal characteristics of the obtained pulses, we demonstrate that the stimulated emissions from the axion cluster with axions sourced by the BH superradiance provide a candidate explanation to the observed fast radio bursts.
\end{abstract}

\maketitle
\newpage

\section{Introduction}
The axion is of particular interest because it 
provides the most credible solution to the strong CP problem of QCD and also becomes one of the popular candidates for dark matter. Due to the spontaneous breaking of the Peccei-Quinn (PQ) symmetry~\cite{Peccei:1977hh}, 
the axion $\phi$ appears as a spin-0 pseudo-Goldstone boson
~\cite{Weinberg:1977ma}. The axion effective field theory reduces its couplings to the Standard-Model particles as follows
\begin{flalign}\label{Leff}
{\cal L}_{\rm eff} = {\alpha_s \over 8\pi f_a}\phi G^a_{\mu\nu} \tilde G^{a\mu\nu} + {\alpha K \over 8\pi f_a}\phi F_{\mu\nu} \tilde F^{\mu\nu} + {1\over f_a} J^\mu \partial_\mu \phi~,
\end{flalign} 
where $\alpha$ and $\alpha_s$ are fine structure coefficients of electromagnetic and strong interactions with $F_{\mu\nu}$ and $G_{\mu\nu}$ as their field strengths, respectively. 
In Eq.~\eqref{Leff}, $\tilde{F}_{\mu\nu}$ and $\tilde{G}_{\mu\nu}$ are the corresponding dual fields for QED and QCD, respectively, while $J^\mu$ denotes the Noether current of the broken PQ symmetry composed of other matter fields, $f_a$ is the decay constant of the axion, and $K$ is a model dependent coefficient of ${\cal O}(1)$. It is seen from Eq.~\eqref{Leff} that all axion effective interactions are characterized by a single energy scale $f_a$ that suppresses the axion couplings to the SM particles. 
Note that the axion models with $f_a$ of the electroweak symmetry breaking scale $v_\text{weak}\sim250$~GeV have been ruled out by the existing experiments~\cite{Barshay:1981ky, Barroso:1981ta}, which favor models namely KSVZ~\cite{Kim:1979if, Shifman:1979if} and DFSZ~\cite{Dine:1981rt, Zhitnitsky:1980tq}, predicting invisible axions with $f_a\gg v_\text{weak}$. Also, the effective field theory~\cite{GrillidiCortona:2015jxo} determines the product of the axion mass $m_a$ and decay constant $f_a$ via
$ m_a   f_a = m_\pi f_\pi = [75.5\text{ MeV}]^2 $, 
where $m_\pi$ and $f_\pi$ are the pion mass and decay constant, respectively. Due to the lightness of the axion, its decay is dominated by the process $a\to \gamma\gamma$ with $\gamma$ representing a photon, leading to the following axion lifetime~\cite{Cheng:1995fd}
\begin{flalign}\label{lifetime}
\tau_a = {256\pi^3 \over K^2\alpha^2 m_a} \left({  f_a \over m_a }\right)^2\,,
\end{flalign} 
where $K$ is a model dependent parameter of ${\cal O}(1)$. In the following, we take $K=1$.

Light bosonic fields such as axions affect the dynamics of rotating black holes through superradiance, which can be used as a probe for fundamental physics. When the Compton wavelength of the axion is of order of the black hole size, it forms gravitational bound states around the black hole, which are much like electron states in a hydrogen. 
{Following Ref.~\cite{Brito:2015oca}, we adopt the notation for the wave function of a gravitational atom as $\Psi_{nlm}$ where $n$, $l$ and $m$ represent the principle, azimuthal, and magnetic quantum numbers. respectively, in analogy to those labeling a hydrogen atom in quantum mechanics.} 
The superradiance instability comes from the complex frequency $\omega$ of the wave function $\Psi_{nlm}\propto e^{-i\omega t}$, which allows $\omega_I>0$ when $\omega_R<m\Omega_H$, where $\Omega_H$ is the angular velocity of the black hole. A positive imaginary part $\omega_I$ of the frequency indicates that the amplitude of the wave function could grow exponentially. 
Let $\mu=Gm_a/\hbar c$ be the reduced mass of the axion and $M$ the mass of the black hole. {It has been well-known that the principle superradiant mode is the one with $(n,l,m)=(2,1,1)$. When the gravitational fine structure constant $\mu M\ll 1$, the real and imaginary parts of its frequency is given by~\cite{Detweiler:1980uk}}
\begin{flalign}\label{frequencies}
\omega_{I211} = \mu\left({a\over M}\right) {(\mu M)^8 \over 24} \,,\quad
\omega_{R211} = \mu - { \mu \over 2 } \left( {M\mu\over 4} \right)^2 \,,
\end{flalign}
where we have kept the leading and subleading order in $\mu M$ for these two parts respectively. The parameter $a$ is defined as $a \equiv J/M$ with $J$ as the angular momentum of the black hole. 
For $\mu M\gg1$, it is shown in Ref.~\cite{Zouros:1979iw} that the scalar field growth is slow and much suppressed. 
Finally, when $\mu M\sim1$, $l=m=1$ is still the fastest growing state, which is demonstrated in Ref.~\cite{Dolan:2007mj} for a rapid rotating black hole with $\mu M\approx0.42$. A comprehensive review of the black hole superradiant instability can be found in Ref.~\cite{Brito:2015oca}. Note that the dominant superradiant mode with $l=m=1$ indicates that the axion distributions are, in general, not spherically symmetrical in the real astrophysical situation.

Owing to axion couplings to photons as given in Eq.~\eqref{Leff}, one remarkable observable effect of the black hole superradiance  is the stimulated emission of photons in the resulting axion cloud. Studies of this phenomenon from coherently oscillating axions were carried out  by Tkachev~\cite{Tkachev:1986tr}, which stipulated spherical symmetry and found that the luminosities of axion clumps were similar to those from quasars. The estimation of the luminosity of an axion cluster was also conducted by Kephart and Weiler in Ref.~\cite{Kephart:1986vc}, which was further developed in Ref.~\cite{Kephart:1994uy}. The spherical symmetry is a key assumption in the latter study, showing that the stimulated photon emission 
could be triggered by a cloud of axions with a high occupation number. More recently, the evolution equations in Ref.~\cite{Kephart:1994uy} of the axion stimulated emission 
were applied to the $2p$ axion cloud formed by the Kerr black hole superradiance~\cite{Rosa:2017ury}, which implicitly made use of spherical symmetry even though the system does not possess such a feature at all. 
{However, in the real situation, with the spherical symmetry broken, the distribution of a axion cloud becomes uneven, and it is generally expected to see new effects to appear, such as the evolution of the shape of the cloud and the photon emitted showing the angular dependence, which demand further investigations.} 
On the other hand, evolution equations of number densities of axions and photons in a non-spherical configuration were given in Ref.~\cite{Chen:2020yvx} but the application of these equations was not fully exploited. Hence, in this paper, we would like to go further by applying the corresponding evolution equations in Ref.~\cite{Chen:2020yvx} to a non-spherical axion configuration which is assumed to be originated from the black hole superradiance. 
We also note that a similar goal has also been pursued in Ref.~\cite{Hertzberg:2018zte}. 
{Here we would like to make some comments on the relationship and differences of our present work with Ref.~\cite{Hertzberg:2018zte}. Note that the non-spherical axion cloud evolution in our work is based on the perturbative calculation of the axion decays and axion-photon interactions, while the calculation in Ref.~\cite{Hertzberg:2018zte} was based on the parametric resonance production of photons from a coherent axion clump, which is a non-perturbative phenomenon. From our point of view, both effects are just two channels for the axion cloud decays so that they should coexist in the system. The situation is much like the relationship between the reheating and preheating processes discussed in \cite{Kofman:1997yn}. When the amplitude of the oscillating axion field is very large and its coupling to photons is sizeable enough, the parametric resonance would become efficient and dictate the axion evolution. However, when the axion-photon coupling is small, or the axion amplitude decreases greatly by depositing its energy into photons, the resonance becomes narrow and even shut off, so that the axion perturbative decays become dominant, which requires to use the Boltzmann equation to describe the evolution. Moreover, when the produced photons are abundant, the processes like backreactions and scatterings by photons would  terminate the resonance and make the axion clump lose the coherence. Due to the extremely tiny coupling of the QCD axions, it is reasonable to expect that the parametric resonance process is always subdominant, so that the correction description of the axion cloud evolution is the associated Boltzmann equations, which is the starting point of our present work.\footnote{Our analysis differs further from Ref.~\cite{Hertzberg:2018zte} in the following several aspects. While Ref.~\cite{Hertzberg:2018zte} considered the effect of the axion clump anisotropy on the production of photons by including the cloud angular momentum, the authors did not consider the evolution of the axion cloud angular distribution. Also, Ref.~\cite{Hertzberg:2018zte} did not take into account the axion production from the black hole superradiance, which acts as a source for axions. Both effects have been thoroughly discussed in the present paper.}}

The paper is organized as follows. We begin by briefly reviewing the relationships among number densities and occupation numbers of axions and photons in the stimulated emission process in Sec.~\ref{SecEq}. Then, we solve these evolution equations in a source-free environment in Sec.~\ref{SecNoSource}, where we also compare the difference between non-spherical and spherically symmetrical setups. In Sec.~\ref{SecSource}, the stimulated emission with the black hole superradiance as an axion source is investigated. We discuss a possible application of our results to the fast radio bursts (FRBs) phenomena in Sec.~\ref{SecDiscussion}, due to the similarities between the predicted photon pulses from non-spherically configured axion clusters and the observed FRB signals. 
Finally, we conclude in Sec.~\ref{SecConc}.

\section{Stimulated emission from an axion cluster}\label{SecEq}
In this section, we present our formalism governing the evolution of the axion and photon number densities in a non-spherical axion cluster. Let us begin by rewriting the Boltzmann equation~\cite{Kephart:1994uy}, which gives the rate of change of the photon number density $n_{\lambda}$ of helicities $\lambda=\pm 1$, 
\begin{flalign}\label{BoltzmannEq}
\frac{dn_{\lambda}}{dt}=&\int dX^{(3)}_\text{LIPS}[f_a(1+f_{1\lambda})(1+f_{2\lambda})-f_{1\lambda}f_{2\lambda}(1+f_a)]|M(a\rightarrow \gamma \gamma)|^2 ~,
\\\nonumber  \text{with }
\int dX^{(3)}_\text{LIPS} =&  \int {d^3p\over (2\pi)^3 2p^0} \int {d^3k_1\over (2\pi)^3 2k_1^0} \int {d^3k_2\over (2\pi)^3 2k_2^0} 
 (2\pi)^4 \delta^{(4)}(p-k_1-k_2) ~.
\end{flalign}
where $f_a$ and $f_{i\lambda}$ are occupation numbers of axions and photons, respectively; $p$, $k_1,2$ are the momentum of the axion and the two photons respectively; $M(a\rightarrow \gamma \gamma)$ denotes the amplitude of an axion decay into a photon pair, which can be derived from the axion-photon interaction in Eq.~\eqref{Leff}. 
For any particle species $i=\gamma$ or $a$ with the occupation number $f_i$, the number density $n_i$ and the total number $N_i$ of these particles are
\begin{flalign}
n_i= \int \frac{d^3p}{8\pi^3}f_i\,, \quad\text{and}\quad N_i = \int_V d^3r\,\, n_i ,
\end{flalign}
respectively.
Expanding the axion and photon occupation numbers in terms of spherical harmonics $Y_{lm}(\Omega)$,
\begin{flalign}\label{DistAP}
f_a(p, r, \Omega, t) =& \sum_{lm}f_{a lm}(t)Y_{lm}(\Omega)  \Theta(p_{\mbox{\tiny max}}-p)\, \Theta(R-r)    \,, \nonumber\\
 f_\lambda(k, r, \Omega, t) =& \sum_{lm}f_{\lambda lm}(t)Y_{lm}(\Omega)   \Theta(R-r) \Theta(k_+-k) \Theta(k-k_-)  \,, 
\end{flalign}
{where we have fixed the axion cluster radius to be $R$ since here we focus on the angular evolution of the cloud. 
$p_{\rm max}$ is the maximal value of the axion momentum, which can be obtained by the escape velocity $\beta = \sqrt{2GM / Rc^2}$ in the cluster of mass $M$. For a given axion with momentum $p$, the outgoing photons can have their momenta within a range of values, depending on the angle of the photon direction with respect to the initial axion. Due to this reason, the yielded photons should have momenta in the range between the minimum and maximum values~\cite{Kephart:1994uy}, denoted as $k_-$ and $k_+$, respectively. We further assumed any values of the axion and photon momenta  appear with equal probability in the allowed ranges at every spatial point, which explains the Heaviside functions involving their momenta $p$ and $k$ in the ansatz of Eq.~\eqref{DistAP}. }

%
%

In the present paper, we focus on the evolution of angular distributions of axions and photons. 
In order to achieve this, we can integrate Eq~.\eqref{BoltzmannEq} over the Lorentz invariant measure $dX^{(3)}_\text{LIPS}$, which can give the following set of equations for axion and photon number densities of different $(l,m)$ modes,
\begin{flalign} \label{BoltzmannEq2}
 \frac{ d n_{\gamma lm}(t) }{ dt }  = &     2 {  n_{a lm}  \over  \tau_a  } + \frac{ 16 \pi^2  }{ \beta m_a^3 \tau_a  } E_{l m }     
   - \frac{ 16 \pi^2  }{ 3 m_a^3  \tau_a } \left(\beta+ {3\over2} \right)   F_{l m }          -  \Gamma_e  n_{\gamma lm}(t)  \,,  \\ \nonumber
\frac{ d n_{a lm}(t) }{ dt }     =&   -  {  n_{a lm}  \over  \tau_a  }  -  \frac{ 8 \pi^2  }{ \beta m_a^3 \tau_a  } E_{l m }      
                                                + \frac{ 8 \pi^2 \beta }{ 3 m_a^3  \tau_a }   F_{l m }  \,,
\end{flalign}
$\Gamma_e \equiv {3c/(2R)}$ is the photon escape rate from the cluster with $c$ denoting the speed of light. The coefficients $E_{lm}$ and $F_{lm}$ are derived from the axion-photon interaction terms and defined by
\begin{flalign} \label{EFlm}
n_a  n_{\gamma}   =&    \sum_{l'm'l'' m''}  n_{a l'm'}  n_{\gamma l'' m''}  Y_{l'm'} Y_{l''m''}    
=   \sum_{lm}  E_{l m } Y_{l m } \, ,  \\ \nonumber
n_{\gamma} ^2  =& \sum_{l'm'l'' m''}  n_{\gamma l'm'}  n_{\gamma l'' m''}  Y_{l'm'} Y_{l''m''} 
=    \sum_{lm}  F_{l m } Y_{l m }   \,.
\end{flalign}
In the following sections, we often use the dimensionless version of Eq.~\eqref{BoltzmannEq2}, in which the time $t$ is normalized in units of the axion lifetime $\tau_a$ so that the escape rate $\Gamma_e$ becomes dimensionless by multiplying $\tau_a$. {We also non-dimensionlized the number densities by using the axion Compton volume defined by ${16\pi^2 / m_a^3 }$. Dividing $n_a$, $n_\gamma$ by this factor recovers real number densities.} 

{Note that we have assumed in Eq.~\eqref{DistAP} flat distributions for the radial and momentum directions. One may wonder to what degree our result would change if we modify such a distribution assumption. As discussed in Ref.~\cite{Chen:2020eer}, the variation of the radial components of axion distribution was shown to only introduce small changes in the final photon signal in the spherically symmetric case. In fact, as long as the angular, radial, time and momentum parts of the distributions for axions and photons can be factorized as in Eq.~\eqref{DistAP}, the integration over the radial and momentum profiles can merely lead to ${\cal O}(1)$ corrections to the constant coefficients in the evolution equations of Eq.~\eqref{BoltzmannEq2}, while the overall structure of these equations is left untouched. From another perspective, the flat distributions in the radial and momentum directions can be viewed as the smearing of the more general distributions. Thus, it is sufficient to implement Heaviside functions for the radial and momentum distributions as in Refs.~\cite{Kephart:1994uy, Rosa:2017ury}.}

\section{source-free non-spherical axion distribution}\label{SecNoSource}

\begin{figure}[ht]
        \centering
\noindent \includegraphics[width=0.5\textwidth]{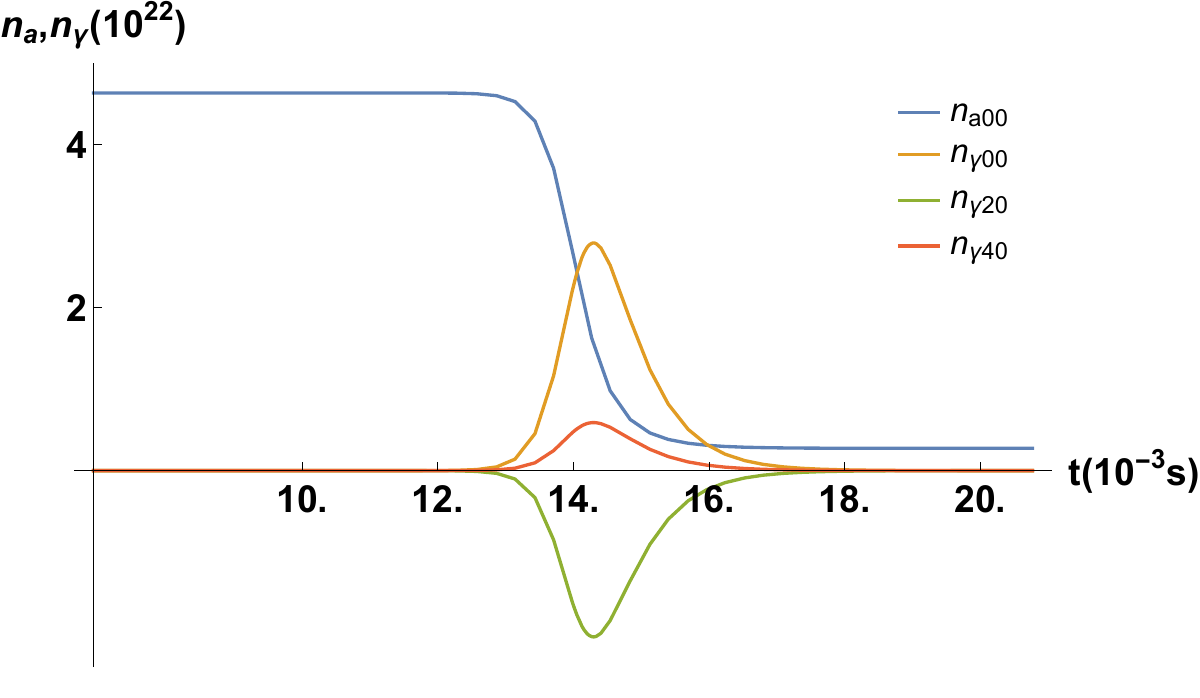}\includegraphics[width=0.5\textwidth]{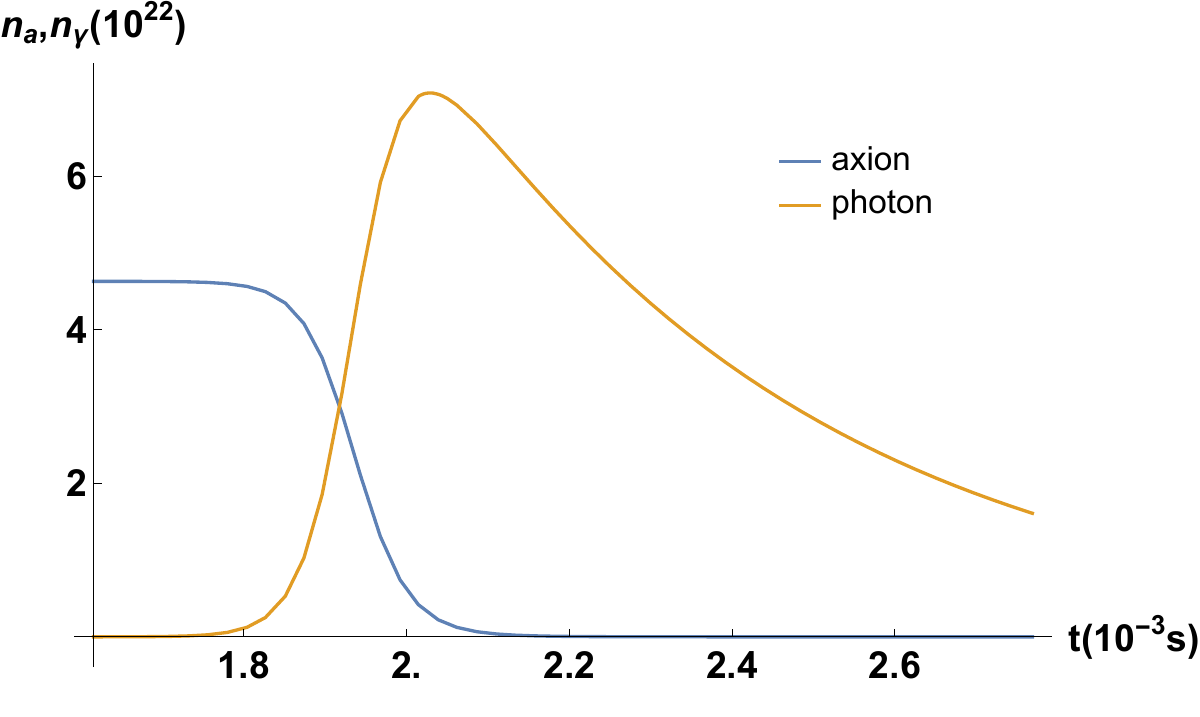}\\
\caption{\label{SF01}
Axion and photon number density components of $n_{a00}$ and ($n_{\gamma00}$, $n_{\gamma20}$, $n_{\gamma40}$) in the non-spherical (left) and spherically symmetrical (right) setups with the same initial axion number density $n_{a00} \sim4.8\times10^{22}$, respectively. The dimensionless densities $n_a$ and $n_\gamma$ are normalized with the inverse of the axion Compton volume ${m_a^3 /16\pi^2 }$.}
\end{figure}

We first consider the axion distribution with $(l,m)=(1,1)$ in the absence of any axion source, {\it i.e.}, without a Kerr black hole in the center of the cluster, so that the amount of axions are fixed.
In this case, the axion wave function is proportional to $Y_1^1$, which translates to the number density of axions $n_a$ being proportional to $Y_1^{1*} Y_1^1 \propto \sin^2 \theta$ with $\theta$ as the polar angle. 
Thus, the distribution of axions can be decomposed in terms of real spherical harmonics $Y_{00}$ and $ Y_{20}$ as follows
\begin{flalign}\label{NaIni}
n_a=& n_{a00}  Y_{00} +  n_{a20}  Y_{20} =  n_{a00}  (Y_{00}-\sqrt{1/5}Y_{20}) \,.
\end{flalign}
It would be interesting to know the distribution of photons released from an axion configuration with the fixed angular dependence as $(l,m)=(1,1)$. Hence, we maintain the shape of the axion configuration  by the relation $n_{a20}  =   -\sqrt{1/5} n_{a00} $ in the present section, leaving the case of an evolving axion angular reliance to the next section. 
Inspired by the axion angular shape, we assume for the moment that the hierarchy of the Boltzmann equations of photons in Eq.~\eqref{BoltzmannEq2} are truncated by
\begin{flalign}
n_\gamma= n_{\gamma00} Y_{00}+ n_{\gamma20} Y_{20} + n_{\gamma40} Y_{40}\,,
\end{flalign}
where we have introduced $n_{\gamma40}$ as a correction besides the $n_{\gamma 00(20)}$ for reasons discussed below. Substituting these into the interaction terms as in Eq.~\eqref{EFlm} can give rise to 
the coefficient $E_{lm}$ and $F_{lm}$ by using the Clebsch-Gordan relations, {\it i.e.}, the product of two spherical harmonics can be decomposed as the sum of these functions. 
Let us take $Y_{20} Y_{20}$ as an example,
\begin{flalign}
Y_{20} Y_{20} = \sqrt{1\over4\pi} Y_{00} + \sqrt{5\over49\pi} Y_{20} + \sqrt{9\over49\pi} Y_{40},
\end{flalign} 
where
\begin{flalign}
Y_{40} ={3\over16}\sqrt{1\over\pi}( 35\cos^4\theta - 30 \cos^2\theta + 3 ),~ Y_{20} ={1\over4}\sqrt{5\over\pi}( 3\cos^2\theta - 1 ),~  Y_{00} =\sqrt{1\over4\pi} ~.
\end{flalign}
In the Boltzmann's equations, the contribution of the stimulated emission to the photon number density comes from the following term
\begin{flalign}
n_a n_\gamma=& 
({ n_{a00}  n_{\gamma00}\over\sqrt{4\pi}} - { n_{a00}  n_{\gamma20}\over\sqrt{20\pi}}  ) Y_{00} 
+ ( { n_{a00}  n_{\gamma20}\over\sqrt{4\pi}} - { n_{a00}  n_{\gamma20}\over\sqrt{49\pi}} - { n_{a00}  n_{\gamma00}\over\sqrt{20\pi}} ) Y_{20} 
-   n_{a00} n_{\gamma20} \sqrt{9\over245\pi} Y_{40} ~,
\end{flalign} 
while
\begin{flalign}
n_\gamma n_\gamma=& 
( {   n_{\gamma00}^2 + n_{\gamma20}^2 \over\sqrt{4\pi}} )Y_{00} + ( { n_{\gamma00}  n_{\gamma20}\over\sqrt\pi} + \sqrt{5\over49\pi} n_{\gamma20}^2 ) Y_{20}
+ \sqrt{9\over49\pi}n_{\gamma20}^2    Y_{40} ~,
\end{flalign} 
accounts for the photon backreaction into axions. By comparing these expressions with Eq.~\eqref{EFlm}, we can extract the relevant coefficients $E_{lm}$ and $F_{lm}$ as follows
\begin{flalign}\nonumber
E_{00} =& { n_{a00}  n_{\gamma00}\over\sqrt{4\pi}} - { n_{a00}  n_{\gamma20}\over\sqrt{20\pi}} \qquad
E_{20} =  { 5 n_{a00}  n_{\gamma20}\over 14\sqrt{ \pi}}  - { n_{a00}  n_{\gamma00}\over\sqrt{20\pi}} \qquad
E_{40} = -   n_{a00} n_{\gamma20} \sqrt{9\over245\pi}
\\
F_{00} =& {   n_{\gamma00}^2 + n_{\gamma20}^2 \over\sqrt{4\pi}}  \qquad
F_{20} =   { n_{\gamma00}  n_{\gamma20}\over\sqrt\pi} + \sqrt{5\over49\pi} n_{\gamma20}^2  \qquad
F_{40} = \sqrt{9\over49\pi}n_{\gamma20}^2
\end{flalign}
By substituting these into the coupled evolution equations in Eq.~\eqref{BoltzmannEq2} already derived in Ref.~\cite{Chen:2020yvx}, we have
\begin{flalign}\label{SFNSS01}
{ dn_{a00} \over dt } =&  - \frac{n_{a00}}{\tau_a} - { 1\over 2\beta } \left( { n_{a00}  n_{\gamma00}\over\sqrt{4\pi}} - { n_{a00}  n_{\gamma20}\over\sqrt{20\pi}} \right) + {\beta\over6} \left( {   n_{\gamma00}^2 + n_{\gamma20}^2 \over\sqrt{4\pi}} \right)\,,
\\\nonumber
{ dn_{\gamma00} \over dt } =& 
\frac{2 n_{a00}}{\tau_a} +  { 1 \over  \beta }  \left( { n_{a00}  n_{\gamma00}\over\sqrt{4\pi}} - { n_{a00}  n_{\gamma20}\over\sqrt{20\pi}} \right)  - \left({\beta\over3}+{1\over2}\right)  {   n_{\gamma00}^2 + n_{\gamma20}^2 \over\sqrt{4\pi}}  - \Gamma_e n_{\gamma00}\,,
\\\nonumber
{ dn_{\gamma20} \over dt } =&
 -{2\over\sqrt5} \frac{n_{a00}}{\tau_a}  +  { 1 \over  \beta }  \left(  { 5 n_{a00}  n_{\gamma20}\over 14\sqrt{ \pi}}  - { n_{a00}  n_{\gamma00}\over\sqrt{20\pi}} \right)  - \left({\beta\over3}+{1\over2}\right) 
\left( { n_{\gamma00}  n_{\gamma20}\over\sqrt\pi} + \sqrt{5\over49\pi} n_{\gamma20}^2 \right)  - \Gamma_e n_{\gamma20} \,,
\\\nonumber
{ dn_{\gamma40} \over dt } =& 
 - {  n_{a00} n_{\gamma20} \over  \beta }   \sqrt{9\over245\pi}  - \left({\beta\over3}+{1\over2}\right) 
\sqrt{9\over49\pi}n_{\gamma20}^2  - \Gamma_e n_{\gamma40} \,,
\end{flalign} 
where the photon escape rate away from the cluster is taken to be $\Gamma_e \tau_a \sim 1.6\times10^{25}$ hereafter. 
If the initial axion cluster has the same mass $M$ as the Earth but with its radius $R$ as $1/10$ of the Earth radius, the maximum velocity of axions can be estimated as $\beta=\sqrt{2GM / Rc^2}\sim10^{-4}$. We have solved the evolution equations in Eqs.~\eqref{SFNSS01} and plot $n_{a00}$, $n_{\gamma00}$, $n_{\gamma20}$ and $n_{\gamma40}$ versus time $t$ in the left panel of Fig.~\ref{SF01} for a hardonic axion~\cite{Kaplan:1985dv} of $m_a=$ 3~eV. For comparison, we also show in the right panel $n_{a00}$ and $n_{\gamma00}$ as functions of time with the same initial axion density except that the axions are distributed spherically.

Here are a few observations. {We have included terms accounting for the axion vacuum decay in Eq.~\eqref{SFNSS01}, which is important in that it is the dominant photon production source at the beginning. Thus, we cannot neglect them. Only when the number of photon density becomes considerable does the stimulated emission takes effect and dominates over the axion decay process. One thing we would like to mention here is that even though the lifetime of an axion is of ${\cal O}(10^{22}\,{\rm s})$, the vacuum decay in the axion cloud could still be effective since such a long lifetime should be balanced by the extraordinarily large axion density of ${\cal O}(10^{22}) (m_a^3/16\pi^2) \approx {\cal O}(10^{35}\,{\rm cm}^{-3})$ for $m_a = 3$\,eV.} The whole decay process is delayed in the non-spherical setup in comparison with that of the spherically symmetrical one. {Technically, this feature can be explained by noticing that there is a cancelation between various interaction terms on the right-hand side of Eqs.~\eqref{SFNSS01}, which leads to the decrease of the rate of the stimulated emission and thus make slow the whole process. From the geometric point of view, when the initial axions is distributed anisotropically as in Eq.~\eqref{NaIni}, even if the average axion density over the cloud is the same as in the spherical case, about half of volumes would suffer the decrease of the axion density, which would need more time for the photons to accumulate to a level so that the stimulated emission becomes efficient. Overall, the process of the a non-spherical configuration would lag behind its spherical counterpart.} 
{Moreover, in the spherically symmetrical setup, the axion density drops to a level far below that of photons quickly, while in the non-spherical case there are still residue axions left in the final state. This can be understood as follows: in the spherically symmetric case, the photon number density increases rapidly to a level so high that the produced photons make stimulated emission efficient. As a result, the axions are depleted to such a low density that they cannot support the stimulated emission any more, which leads to the decrease of the photon density in the cloud. In contrast, for the non-spherical setup, it is shown in the left plot in Fig.~\ref{SF01} that photons of different angular momentum modes do not reach high densities. Even at the peak region, these photon densities do not exceed the corresponding axion one, so that the stimulated emission stops when the photon density cannot source the process. In the end, some amount of axions remain. }


\begin{figure}[ht]
        \centering
\noindent\includegraphics[width=0.5\textwidth]{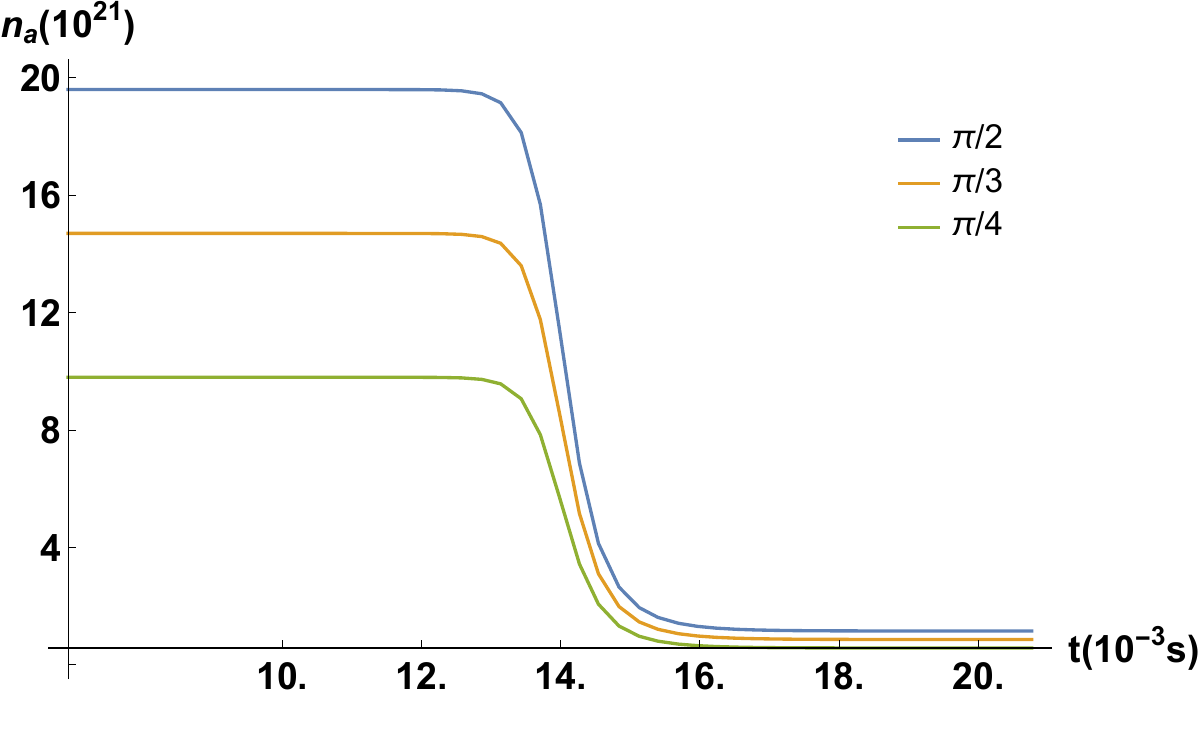}\includegraphics[width=0.5\textwidth]{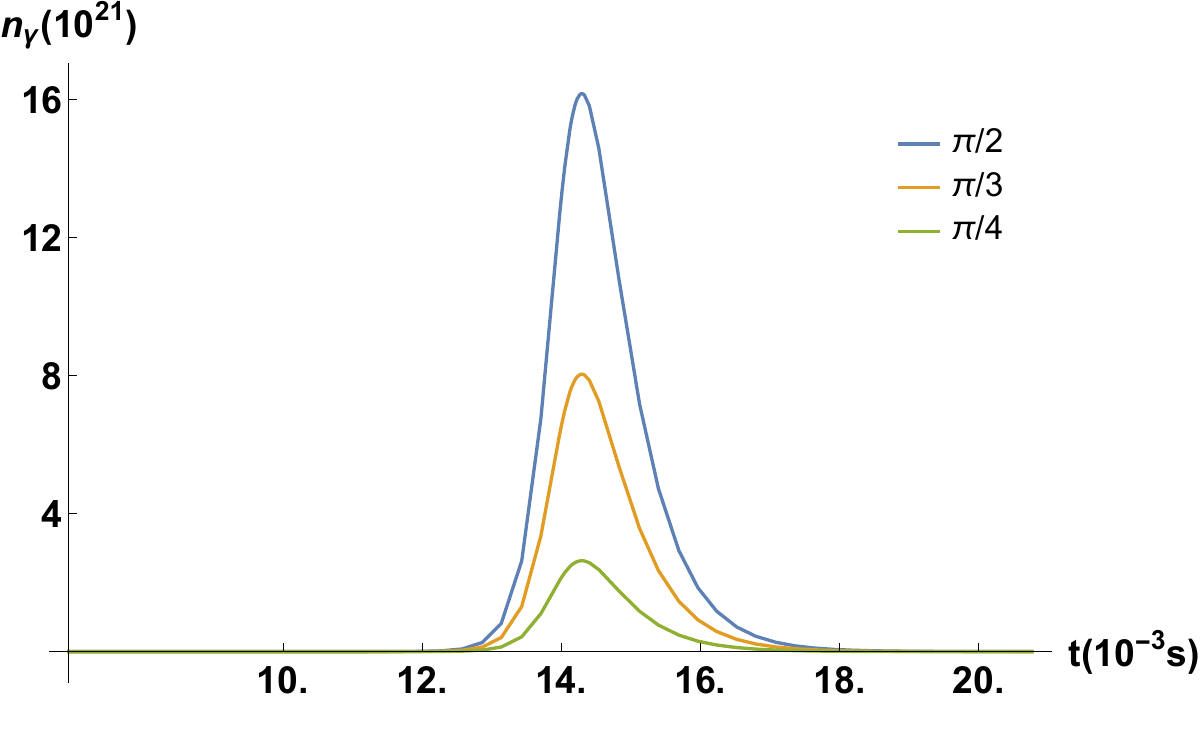}
\caption{\label{SF02}
The axion (left) and photon (right) number densities evolve as a function of time $t$ at $\theta=\pi/2$ (blue), $\pi/3$ (brown) and $\pi/4$ (green), respectively. 
The number densities $n_a$ and $n_\gamma$ are non-dimensionized using the inverse of the axion Compton volume ${m_a^3 /16\pi^2}$. Closer to the equatorial plane ($\theta=\pi/2$), the stimulated emission is stronger since more axions are concentrated there. In contrast, near the polar area ($\theta\sim0$) where axions become dilute, the photon production is very weak. 
}
\end{figure}
In Fig.~\ref{SF02}, we also plot the total number densities of axions $n_a$ (left panel) and photons $n_\gamma$ (right panel) as functions of time $t$ for different azimuthal angles $\theta=\pi/2$, $\pi/3$ and $\pi/4$, respectively. It is shown in the right panel of Fig.~\ref{SF02} that the stimulated emission is strongest on the equatorial plane with $\theta=\pi/2$ and it diminishes as $\theta$ decreases. Close to the polar area where $\theta\approx 0$ or $\pi$, this effect does not manifest itself, which can be understood by the angular distribution of axions in the mode of $(l,m)=(1,1)$. Our plots in Figs.~\ref{SF01} and \ref{SF02} illustrate the general features of the stimulated emission for a non-spherical axion cluster without axion sources. 

\begin{figure}[ht]
        \centering
\noindent \includegraphics[width=0.5\textwidth]{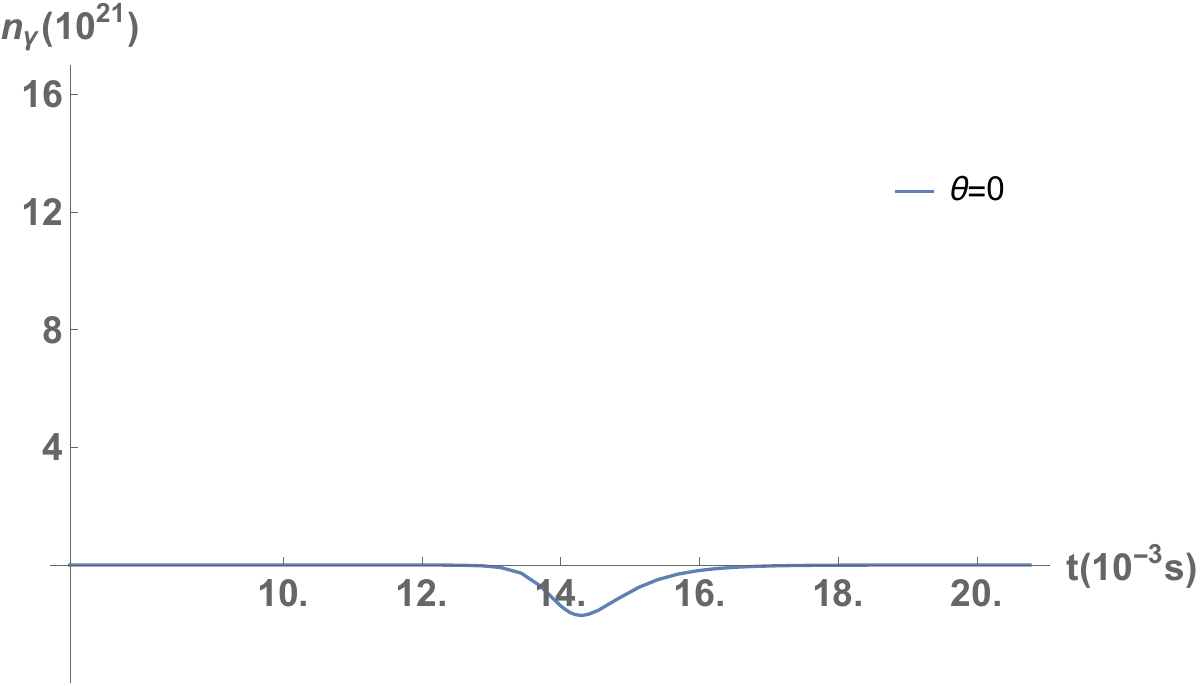}\includegraphics[width=0.5\textwidth]{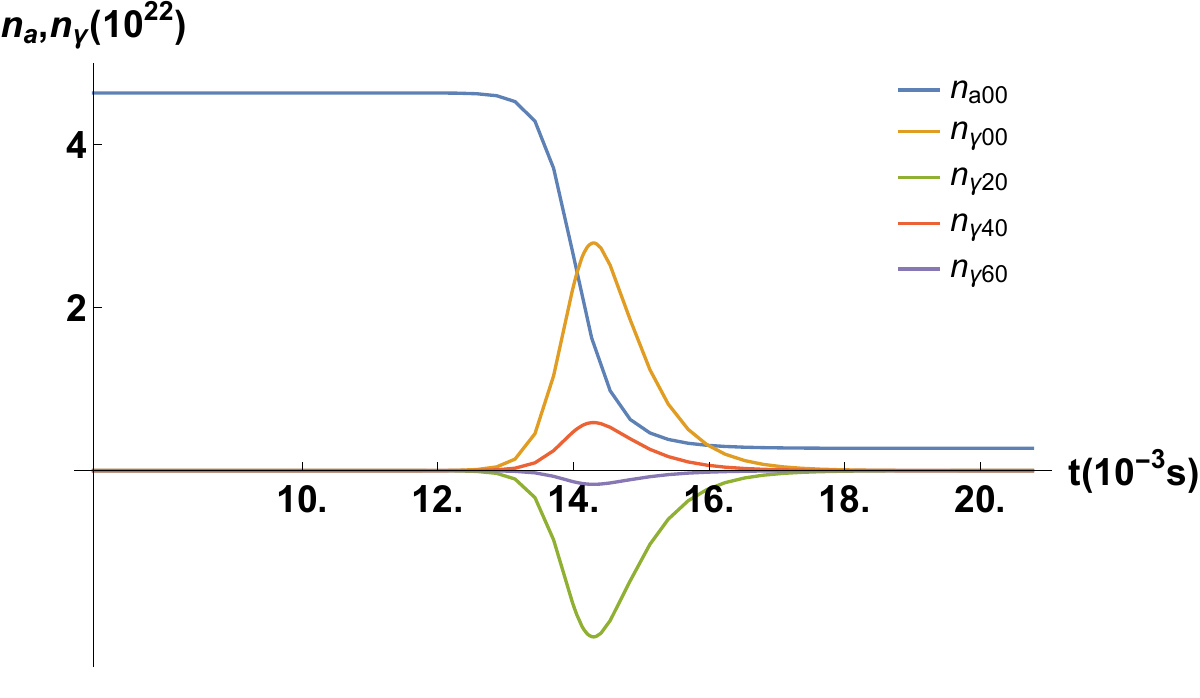}
\caption{\label{SF03}
The left panel demonstrates the evolution of the total photon density $n_\gamma$ at $\theta=0$ by introducing only components $n_{\gamma00}$ and $n_{\gamma20}$, while
the right panel shows the effect of one higher photon mode $n_{\gamma60}$ compared with other lower angular modes. }
\end{figure}

Finally, we would like to point out the importance of the inclusion of the photon component $n_{\gamma40}$ in our numerical solutions, even though there is no direct spontaneous emissions from axions to feed into it. 
{In order to show this point, we solve the evolution equations in Eq.~\eqref{SFNSS01} again but this time we do not consider the photon density component $n_{\gamma40}$ and its associated equation. The final result is displayed in the left panel of Fig.~\ref{SF03}, where we show the evolution of the total photon density around the polar region with $\theta \approx 0$. We observe a trough at $t \simeq 14 \times 10^{-3}$\,s, representing a negative photon density, which is physically unacceptable. This signals that the whole system truncated at $n_{\gamma20}$ is not complete, and other components such as $n_{\gamma40}$ should be included in the computation. In contrast, as discussed before, when $n_{\gamma40}$ and its associated Boltzmann equations are included as in Eq.~\eqref{SFNSS01}, the trough around the polar region $\theta \approx 0$ disappears and the total photon density is always nonzero at any azimuthal angles. This indicates that the inclusion of the component $n_{\gamma40}$ is necessary to obtain a consistent and complete description of the stimulated emission in the non-spherical situations.} {Moreover, one might wonder if it is enough to truncate our Boltzmann hierarchy at the mode $n_{\gamma40}$. In order to investigate this problem, we rerun our simulation again by including the next higher photon mode $n_{\gamma60}$. The result is shown in the right panel of Fig.~\ref{SF03}, where the higher mode $n_{\gamma60}$ is seen to be smaller than other lower angular modes by about one order of magnitude. Therefore, this higher angular mode gives little effects on the evolution of the system and thus can be neglected. As for axions, since the main source is from the black hole superradiance, which is in the $(l,m)=(1,1)$ mode, so that it is enough to truncate the axion density hierarchy at the $(2,0)$ mode.}

\section{Stimulated Emission of a Non-spherical Axion Distribution with Axion Source from Superradiance}\label{SecSource}
\begin{figure}[ht]
        \centering
\includegraphics[width=0.50\textwidth]{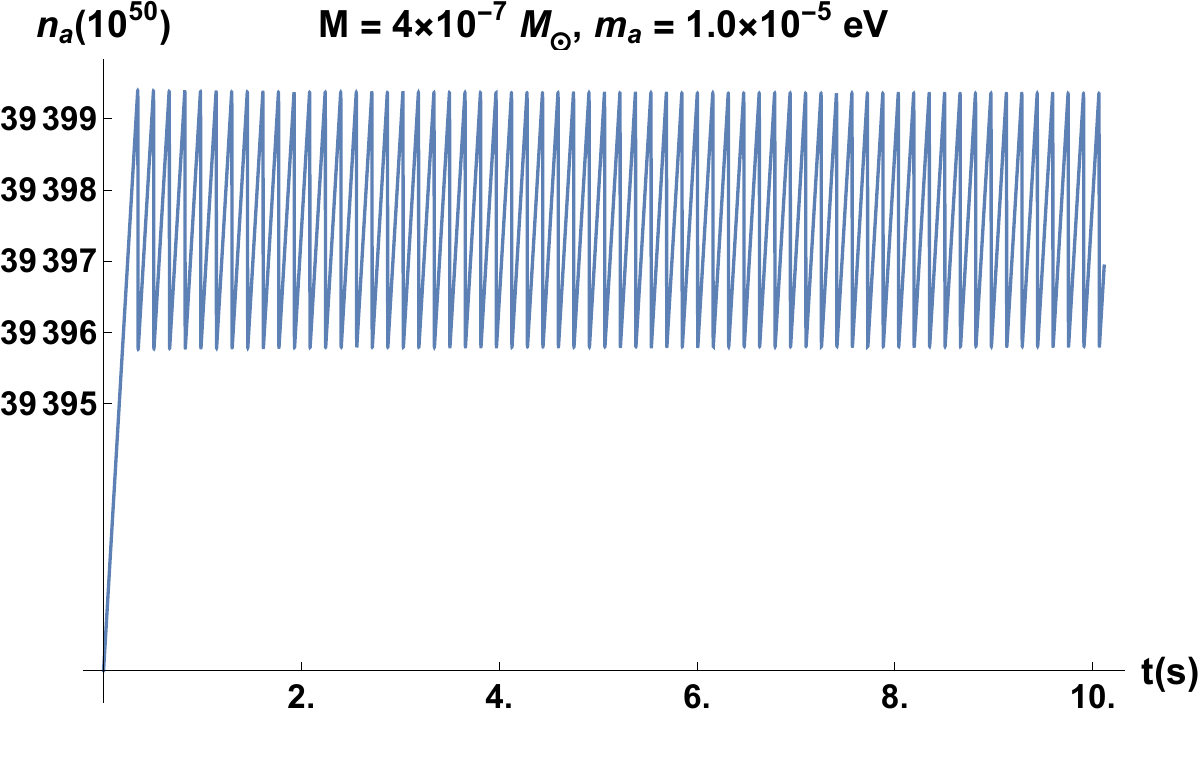}\includegraphics[width=0.50\textwidth]{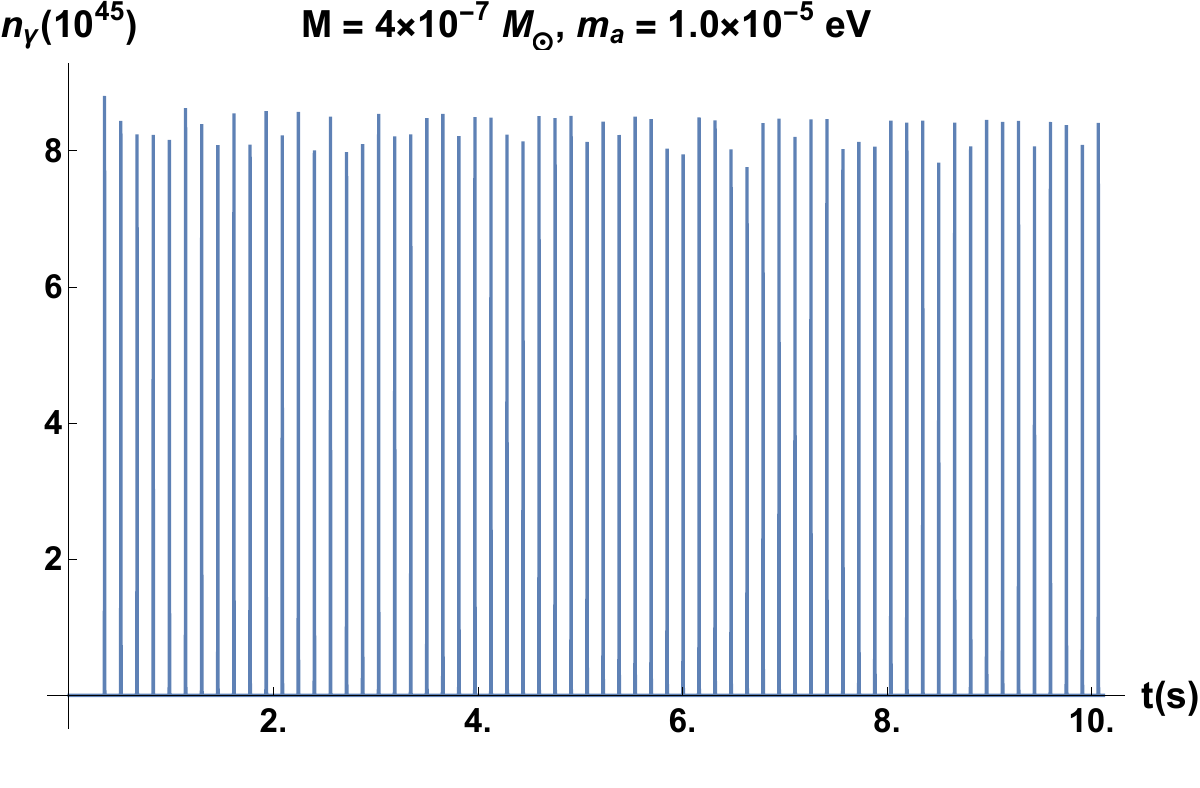}
\caption{\label{SphSym}
Similar to Fig.~2 in Ref.~\cite{Rosa:2017ury} but with different initial conditions of the axion and black hole masses as $m_a = 10^{-5}$~eV and $M=8\times10^{23}$~kg $\sim4\times10^{-7}M_\odot$, respectively. The left (right) panel shows evolution of the axion (photon) number density in a spherically symmetric system.}
\end{figure}

In the last section, we have only considered the stimulated emission in the case with a source-free axion cluster. On the other hand, if there is a highly rotating black hole in the center of the cluster, it is generally expected that axions can be continuously produced via the superradiance effect. In other words, we should consider the stimulated emission in presence of the axion source from the black hole superradiance. Note that this situation has already been studied in Ref.~\cite{Rosa:2017ury}. {In Fig.~\ref{SphSym}, we show similar evolutions of number densities of axions (left panel) and photons(right panel) as those given in Ref.~\cite{Rosa:2017ury}, where the axion and black hole masses are taken to be $m_a=10^{-5}$ eV and $M=8\times10^{23}\,{\rm kg} \approx 4\times10^{-7}M_\odot$, respectively. The timing, shape and amplitude of the resultant axion density and photon pulses in these plots are almost identical to those presented in Ref.~\cite{Rosa:2017ury}, but with different initial conditions. We have also checked that the conditions for black hole superradiance in Eqs.~\eqref{frequencies} are satisfied as $\mu M\sim0.03\ll1$ and $\omega_R\sim 0.15\Omega_H<m\Omega_H$.}
It is evident from Fig.~\ref{SphSym} that the photon pulses from the axion stimulated emission are uniformly repeated in both pace and amplitude. Due to the similarities of the yielded photon emission with the FRB spectra, Ref.~\cite{Rosa:2017ury} has also pointed out the possible explanation of the mysterious FRBs as the stimulated emission in the black-hole sourced axion clusters. {Here one may wonder why the final signal are photon pulses, rather than being in a steady-state configuration in which axion and photon pulses approach constant values. Note that the latter state can be achieved only when the rates of superradiant growth of axions, the photon stimulated emission, and the photon escape from the clump are all balanced delicately in the system. Obviously, such a state requires the model parameters to be considerably fine-tuned, so that it is highly impossible to realize it. On the contrary, if the three relevant rates do not perfectly match, the pulse signal would be more probable. Let us consider the case in which the escape time is longer than those for superradiance and stimulated emission. Then axions and photons would accumulate due to the latter two effects within the escape time scale when the photons would be released into the space outside the axion cloud, which gives rise to the pulse signals. The whole process would repeat again and again, which produces the consecutive photon pulses shown in Fig.~\ref{SphSym}.}

We would like to mention that the above spectra are obtained from the coupled differential evolution equations derived in Ref.~\cite{Kephart:1994uy}, where the spherical symmetry has been implicitly used. However, it is well-known that the axions from a Kerr black hole superradiance should be dominantly distributed in the $(l,m)=(1,1)$ mode, which is obviously non-spherical. Thus, it is necessary to revisit the stimulated emission with axions sourced by the black hole superradiance again by applying Eq.~\eqref{SFNSS01} for a genuine non-spherical setup, which is the main goal of the present section.

For completeness, we rewrite the basic evolution equations given in Ref.~\cite{Chen:2020yvx} for the superradiance-lasing mechanism in a non-spherical cluster and modified by including an axion source term accounting for the axion exponential growth from the black hole superradiance
\begin{flalign}\label{NSSEvoEqua}
{ dn_{a00} \over dt } =& \Gamma_n n_{a00} - \frac{n_{a00}}{\tau_a} - { E_{00}\over 2\beta } + {\beta\over6}F_{00},  \qquad
{ dn_{\gamma00} \over dt } =  \frac{2n_{a00}}{\tau_a} + { E_{00}\over  \beta } - ({\beta\over3}+{1\over2})  F_{00} - \Gamma_e n_{\gamma00}
\nonumber\\
{ dn_{a20} \over dt } =& \Gamma_n n_{a20} - \frac{n_{a20}}{\tau_a} - { E_{20}\over 2\beta } + {\beta\over6}F_{20},  \qquad
{ dn_{\gamma20} \over dt } = \frac{2 n_{a20}}{\tau_a} + { E_{20}\over  \beta } - \left({\beta\over3}+{1\over2}\right)  F_{20} - \Gamma_e n_{\gamma20}
\nonumber\\
{ dn_{a40} \over dt } =&  - \frac{n_{a40}}{\tau_a} - { E_{40}\over 2\beta } + {\beta\over6}F_{40},  \qquad
{ dn_{\gamma40} \over dt } = \frac{2 n_{a40}}{\tau_a} + { E_{40}\over  \beta } - ({\beta\over3}+{1\over2})  F_{40} - \Gamma_e n_{\gamma40}\,.
\end{flalign}
where $E_{lm}$ and $F_{lm}$ are coefficients of the modes $Y_{lm}$ in $n_a n_\gamma$ and $n_\gamma^2$, respectively, and $\Gamma_n$ is the growth rate induced by the imaginary part of the frequency in Eq.~\eqref{frequencies}. We still need to extract $E_{lm}$ and $F_{lm}$ from the products $n_a n_\gamma$ and $n_\gamma^2$, given by
\begin{flalign}
n_a n_\gamma =&  E_{00} Y_{00}+ E_{20} Y_{20}+ E_{40} Y_{40},
\qquad
n_\gamma^2 = F_{00} Y_{00}+ F_{20} Y_{20}+ F_{40} Y_{40},
\end{flalign}
respectively. The total axion and photon number densities $n_a$ and $n_\gamma$ can be decomposed into the spherical harmonic components as follows
\begin{flalign}
n_a= n_{a00} Y_{00}+ n_{a20} Y_{20}+ n_{a40} Y_{40}\,,
\qquad
n_\gamma= n_{\gamma00} Y_{00}+ n_{\gamma20} Y_{20}+ n_{\gamma40} Y_{40}\,,
\end{flalign}
respectively.
According to the general argument~\cite{Brito:2015oca},
the axion wave function arising from the black hole superradiance should be dominated by the $(l,m)=(1,1)$ mode proportional to $Y_1^{\pm1}$, which results in the initial number density of the axion depending on $Y_1^{\pm1*}Y_1^{\pm1}\propto \sin^2\theta\propto Y_{00}-\sqrt{1/5}Y_{20}$. 
As found in the previous section, we also include $n_{a(\gamma)40}$ for completeness of our computation. 

\begin{figure}[h!]
        \centering
\includegraphics[width=0.5\textwidth]{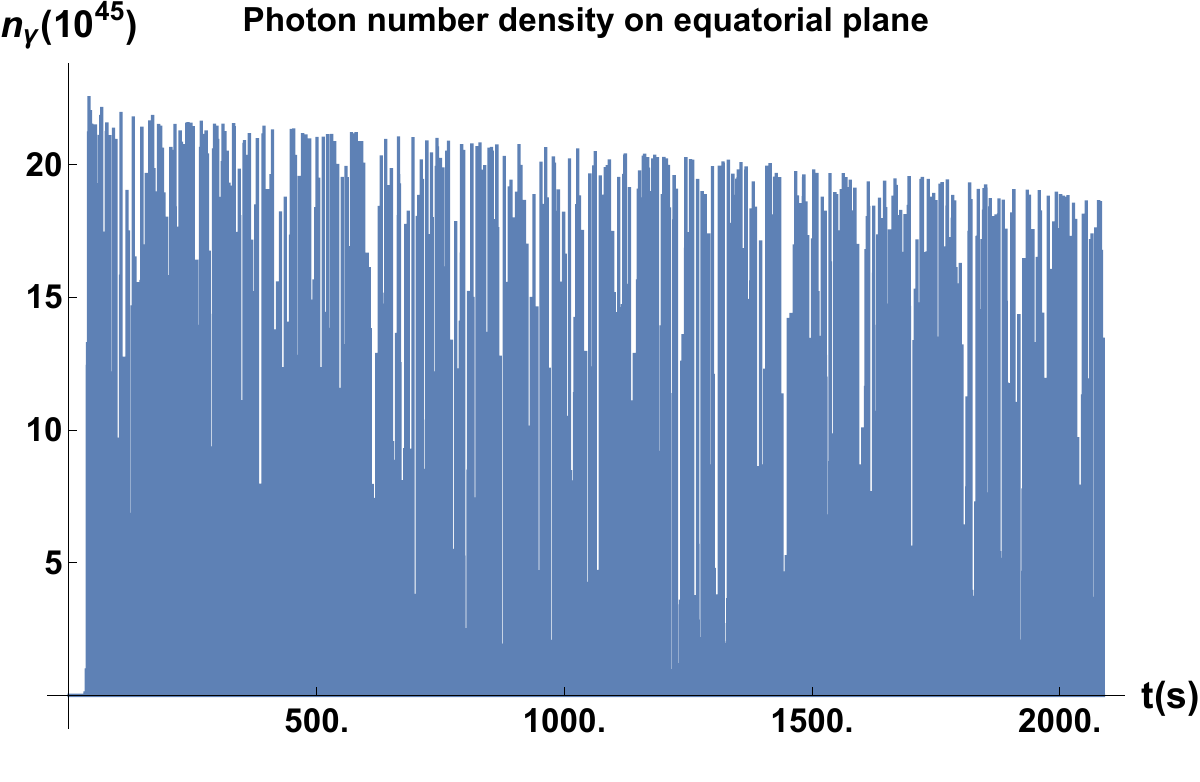}\includegraphics[width=0.5\textwidth]{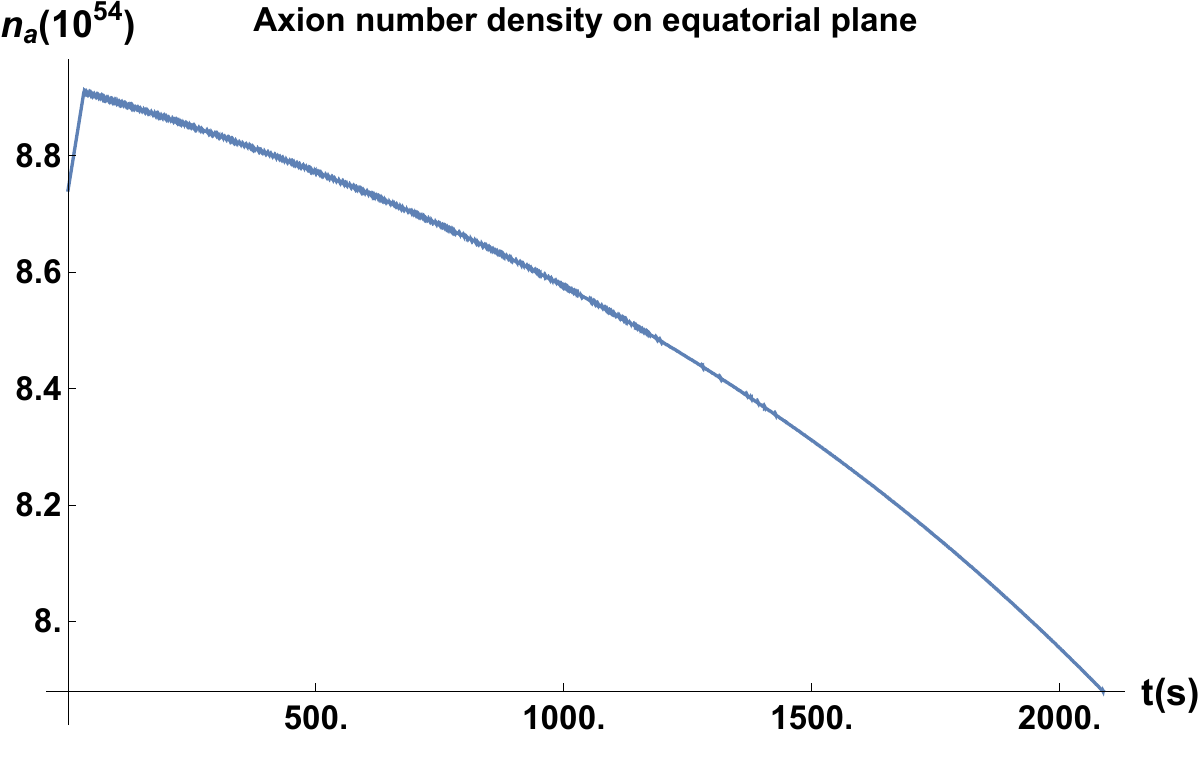}
\\
\includegraphics[width=0.5\textwidth]{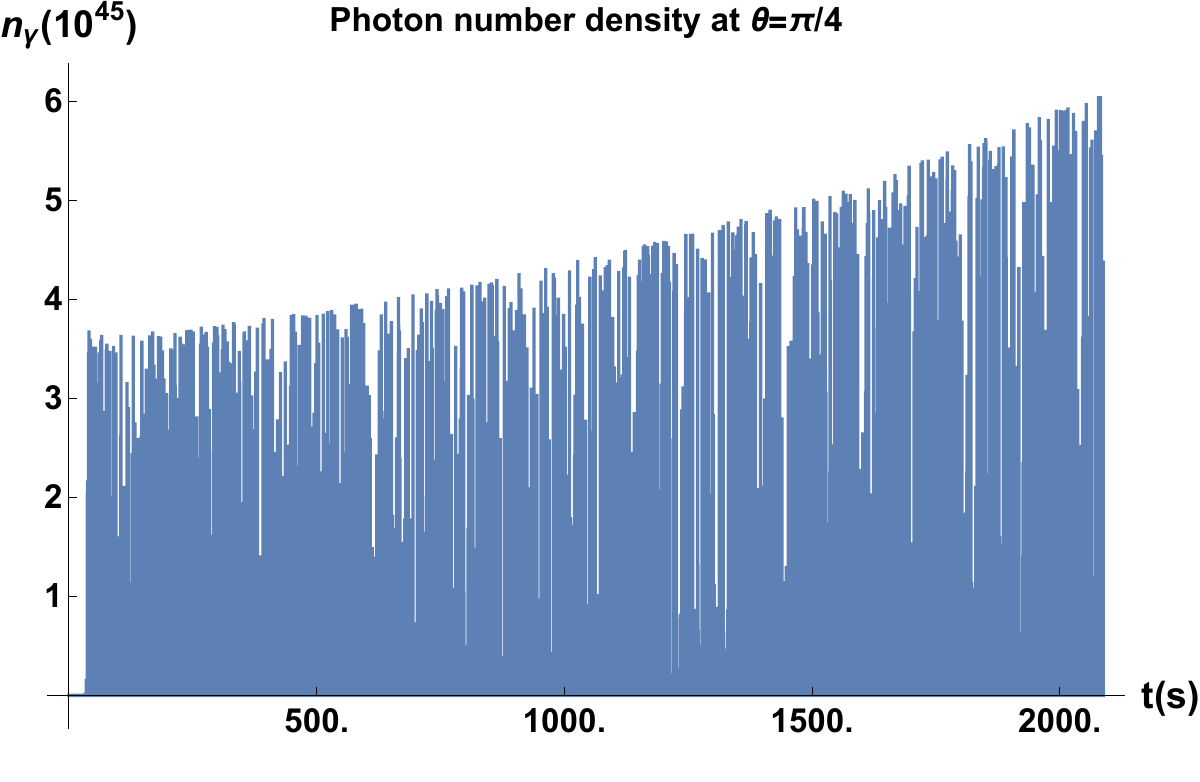}\includegraphics[width=0.5\textwidth]{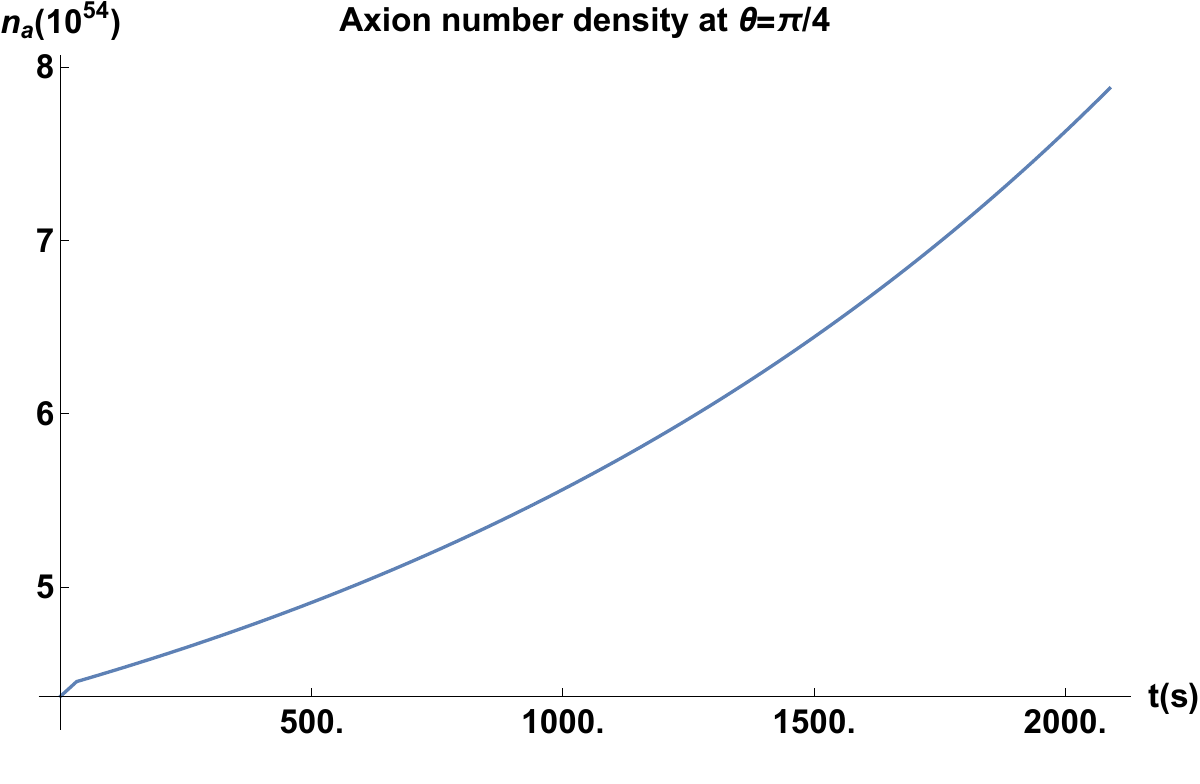}
\\
\includegraphics[width=0.5\textwidth]{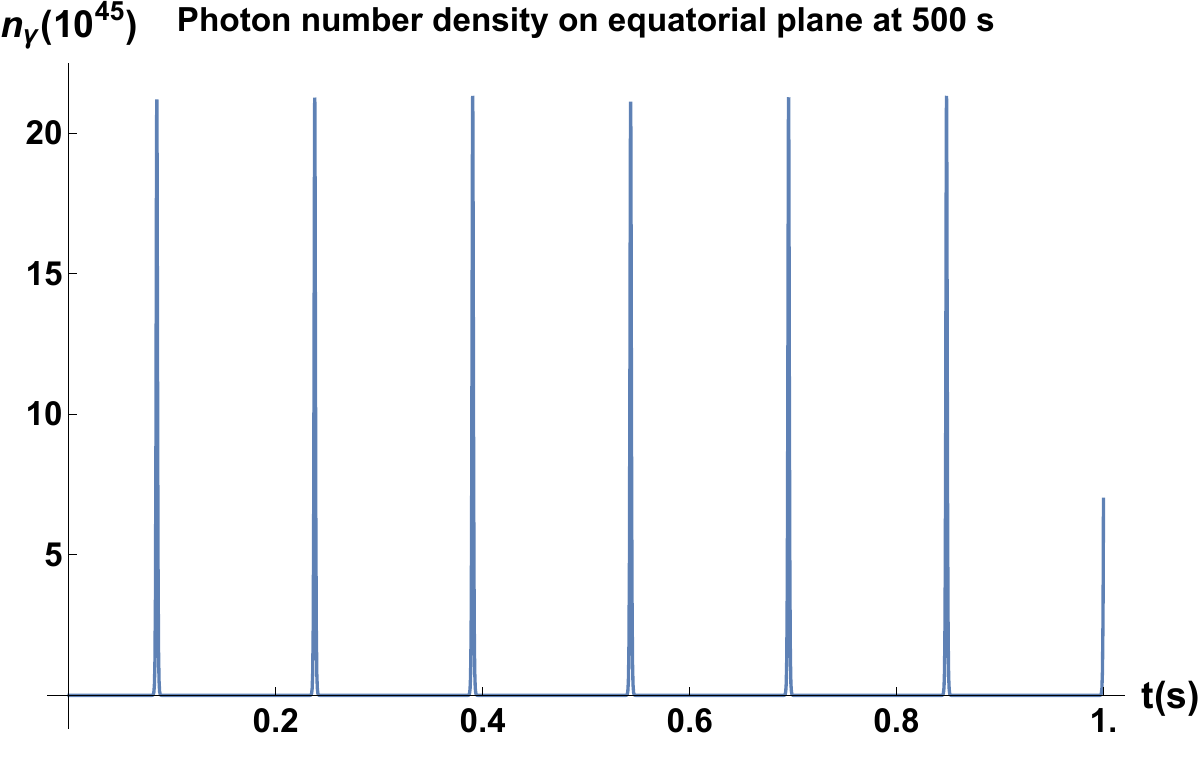}\includegraphics[width=0.5\textwidth]{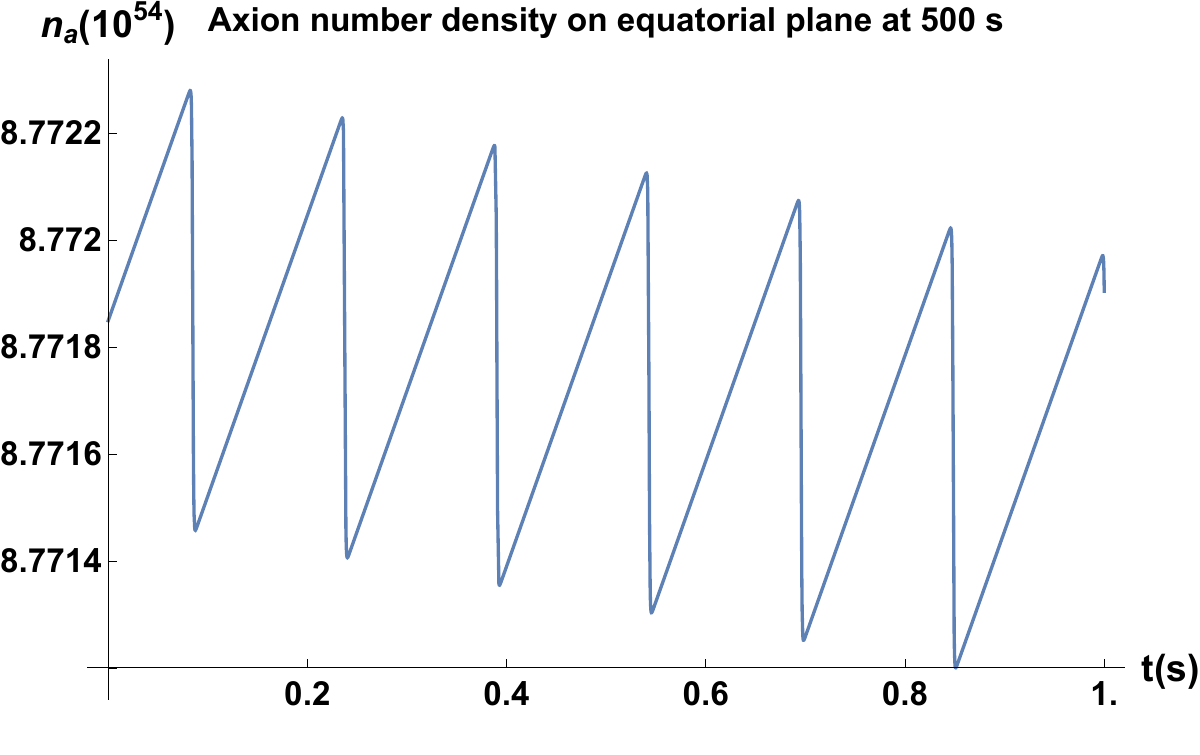}
\caption{\label{NSphSym}
Numerical evolution of the photon (left panels) and axion (right panels) densities as functions of time $t$ on the equatorial plane of $\theta=\pi/2$ (top panels) and at $\theta = \pi/4$ (middle panels). On the bottom two plot we show a zoom-in view of photon and axion number densities on the equatorial plane around $t = 500$\,s.}
\end{figure}

By solving differential equations in Eq.~\eqref{NSSEvoEqua}, we can yield the evolution of number densities of photons and axions in the direction of the polar angle $\theta=\pi/2$ and $\theta=\pi/4$ as displayed in Fig.~\ref{NSphSym}. On the equatorial plane with $\theta=\pi/2$, {as more axions are produced in the (1,1) state by the black hole superradiance, the axion density increases at the beginning. The stimulated emission rate also becomes larger with the accumulation of emitted photons. As the photon density continues to grow, the depletion of axions from stimulated emissions becomes faster than the axion supply from the superradiance, then the axion number density in the cluster would reach its maximal value and starts deceasing, which explains the abrupt jump behavior shown in the top-right plot of Fig.~\ref{NSphSym}.} In comparison, at $\theta=\pi/4$, the axion number density keeps increasing because the production of axions from the superradiance is faster than the loss due to the stimulated emission. The reason is that there are not enough photons created along this direction to deplete axions fast enough. {By continuing the trends of axion densities at these two polar angles, we expect that these two curves would meet in the large $t$ limit. As proved numerically later, this expectation is true. In fact, the axion densities of all directions would coincide exactly at one moment, making the axion density distributed spherically symmetric. } 


{During this redistribution of the axion number density, photon pulses are released with its evolving amplitudes, seen on the left three plots in Fig.~\ref{NSphSym}. The pulses are sharp but regularly distributed in time. On the equatorial plane where $\theta = \pi/2$, the magnitudes of photon pulses decrease gradually as the axion density drops down in most times. At $\theta=\pi/4$, the increasing magnitudes of photon pulses also follow the same trend of the axion density.}

\begin{figure}[ht]
        \centering
\includegraphics[width=0.50\textwidth]{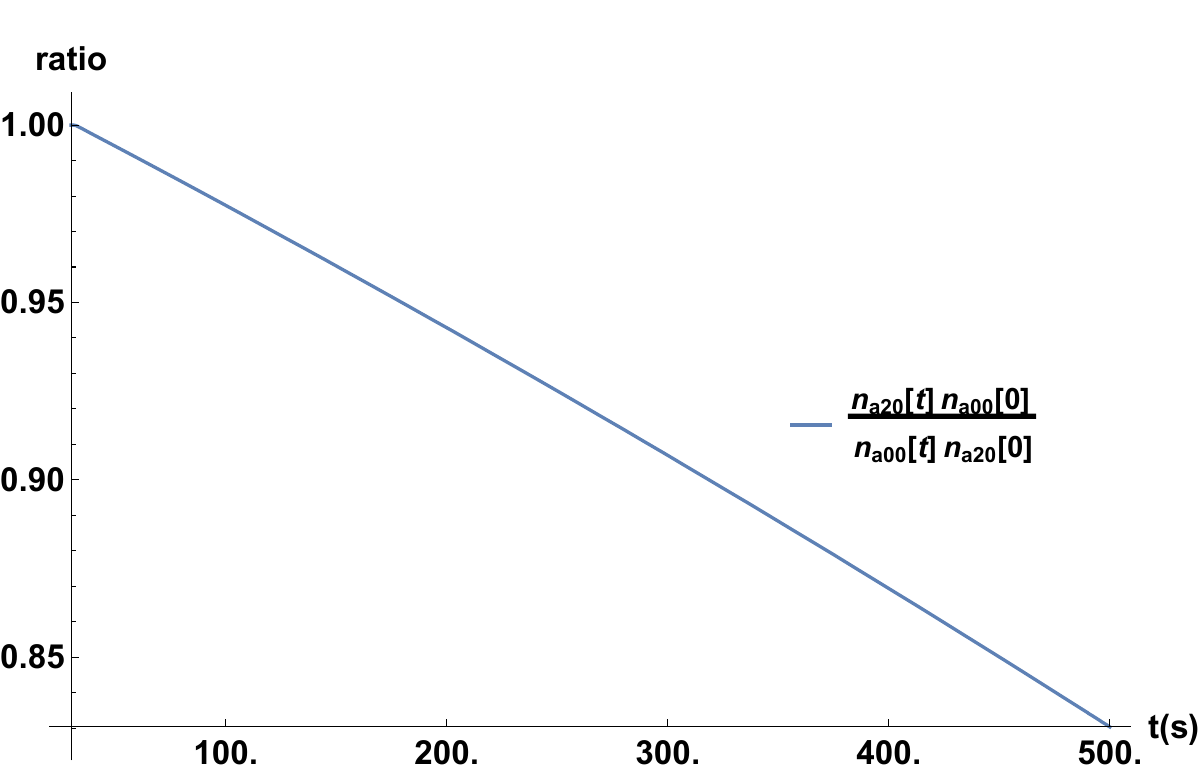}
\caption{\label{deviate11}
The ratio $n_{a20}(t)n_{a00}(0)/(n_{a00}(t) a_{a20}(0))$ as a function of time.}
\end{figure}
\begin{figure}[ht]
        \centering
\includegraphics[width=0.80\textwidth]{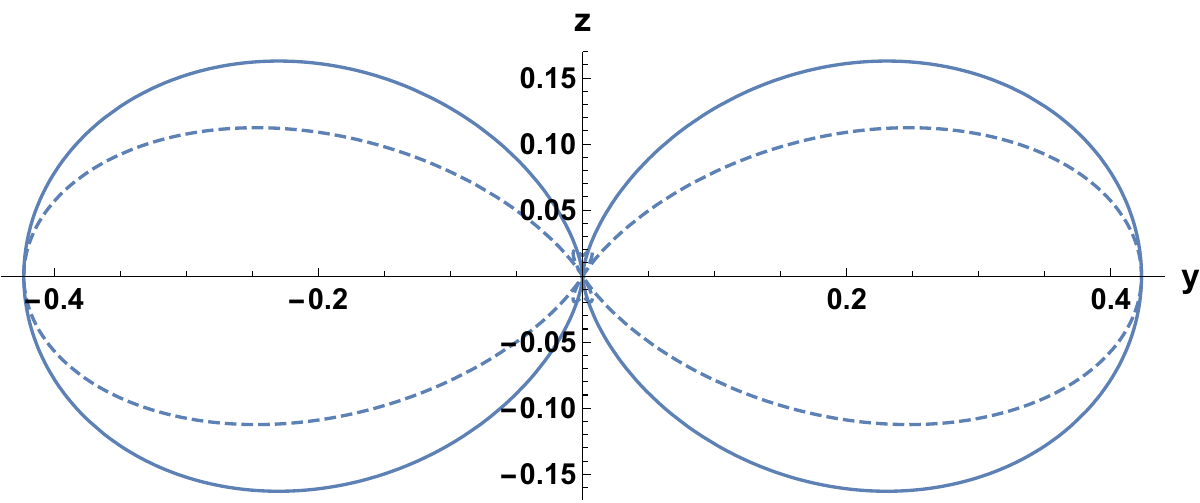}
\caption{\label{apclouds}
Comparison of the cross-sections in the $z-y$ plane of the angular distribution of axions (solid) $Y_{00}+(n_{a20}/n_{a00})Y_{20}$ and photons (dashed) $Y_{00}+(n_{\gamma20}/n_{\gamma00})Y_{20}$ where $n_{a20}/n_{a00} = - 1/\sqrt{5}$ and $n_{\gamma20}/n_{\gamma00}=-0.75$. The tick marks tell the aspect ratio of the shape of the cloud.} 
\end{figure}

The question of whether the axion cloud in the growing mode $(l,m)=(1,1)$ is able to maintain its ``eigenstate'' shape is not discussed in previous works in which the evolution process is approximated by fixing the spherical shape of the cluster. Here, our initial condition is set in the mode $(l,m)=(1,1)$ by the relation $n_{a20}(0)=-\sqrt{1/5}n_{a00}(0)$. 
It turns out that, once the stimulated emission becomes important, the shape of the axion cloud may evolve and cannot stay in the eigenstate $(l,m)=(1,1)$. This feature becomes more transparent in Fig.~\ref{deviate11} that the ratio $n_{a20}/n_{a00}$ starts to deviate away from $-\sqrt{1/5}$. 
The angular distributions of photons are noticeably different from those of axions as shown in Figs.~\ref{deviate11} and \ref{apclouds}. Especially, the ratio $n_{\gamma20}/n_{\gamma00}$ becomes smaller as the system evolves, and deviates more from the $(l,m)=(1,1)$ distribution. 
The above observation can be understood as follows. Initially, the axion cloud is in the $2p$ mode with $(n,l,m)=(2,1,1)$, in which the axion density concentrates on the equatorial plane, whereas there are much less axions along the polar direction. The evolution induced by the spontaneous and stimulated emissions gradually moderates this difference by decreasing $|n_{a20}/n_{a00}|$. 
On the other hand, the photon distribution in the cloud skews towards the equatorial plane, which is compared to that of the axion. The skewness of the axion distribution is directly related to its initial $(l,m)=(1,1)$ wave function, while the photon shape is caused by the non-linear nature of the stimulated decay. The initiation of the stimulated emission requires quite a large density of axions. Therefore, in the direction of higher latitude, the axion cloud becomes less dense so that it cannot make the stimulated emission significant. This results in fewer photons produced at high latitude, which explains the skewness of the photon angular shape. 
We have displayed in Fig.~\ref{apclouds} the typical cross sections of $Y_{00}+(n_{a20}/n_{a00})Y_{20}$ and $Y_{00}+(n_{\gamma20}/n_{\gamma00})Y_{20}$ in polar coordinates when $n_{\gamma20}/n_{\gamma00} = -0.75$.

By observing from the right two panels of Fig.~\ref{NSphSym} that the axion density on the equatorial plane decreases while $n_a$ at $\theta=\pi/4$ increases in the time period of the simulation, we expect that these two curves should coincide at some late time, which means that the axion densities at these two directions will then become the same. In order to prove this expectation, we solve the differential equations in Eq.~\eqref{NSSEvoEqua} again but with a longer time period and at more directions with $\theta=$ $\pi/2$, $2\pi/5$, $\pi/3$, $\pi/4$, $\pi/6$, $\pi/8$, and $0$, respectively. 
\begin{figure}[th]
        \centering
\includegraphics[width=0.5\textwidth]{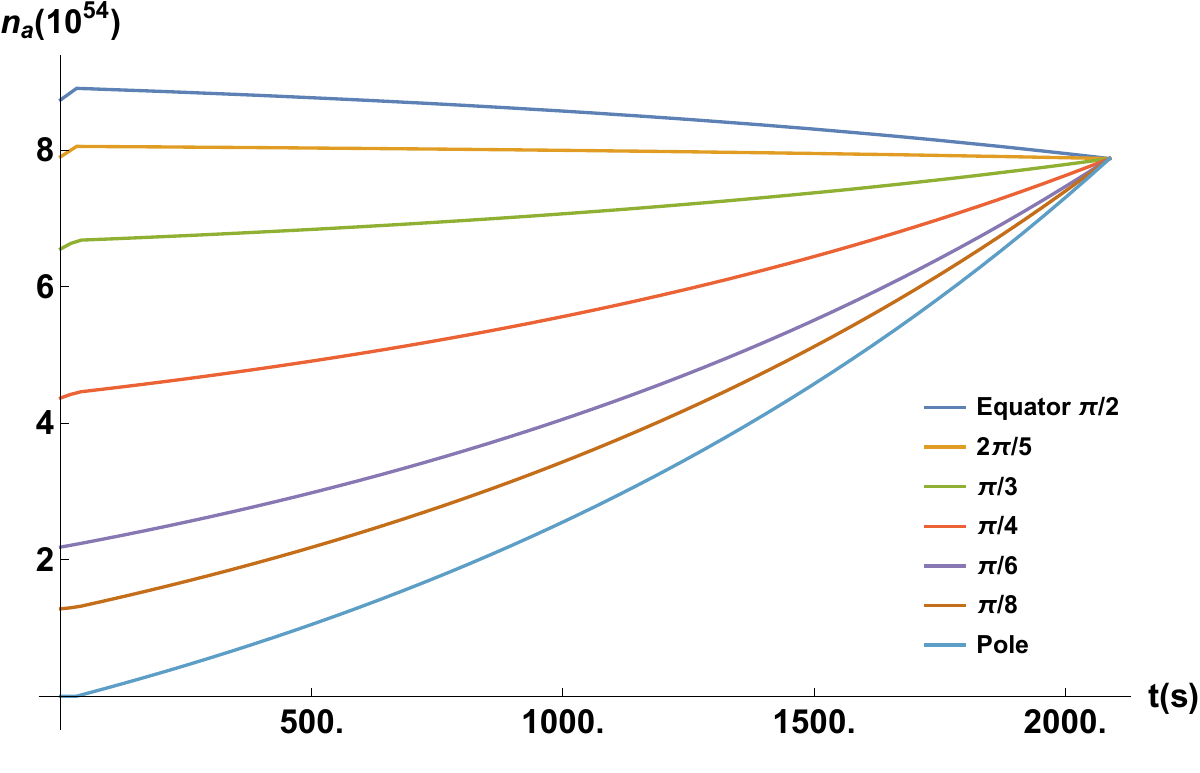}
\caption{\label{uniform}
Uniformization of a non-spherical axion cluster initially of the wave function mode $(l,m)=(1,1)$.}
\end{figure}
The result is shown in Fig.~\ref{uniform}. Indeed, all curves that represent the axion densities in different directions intersect exactly at the same magnitude simultaneously, which infers that the distribution of axions becomes spherically symmetrical. We can explain this result by noticing the fact that, beginning at the $l=m=1$ state, the loss of axions from the stimulated emission overcomes the superradiant growth in the equatorial region with initial high axion densities while the opposite takes place in the polar region with low axion densities. In the end, the interplay of the stimulated emission and the black hole superradiance will reach a balance in which the entire axion cloud approaches a spherical distribution. 
{Here we would like to emphasize that the sphericalization process has little to do with the angular momenta or the movement states of axion particles in the cloud, which can be seen in Eq.~\eqref{DistAP} that the momenta of axions and photons are assumed to only have radial components and cannot give rise to angular momenta in our setup. Because this phenomenon is simple and generic, we believe that the conclusion would be valid even when we consider non-trivial angular momentum distributions in the axion cloud. }

Due to the generality of the previous discussion, it is expected that this conclusion is still valid when performing more precise simulations by including the subdominant higher multipole modes like $n_{\gamma60}$ or $n_{a40}$.  After the axion cloud becomes spherical, the system is reduced to a quasi-stable profile analyzed in Ref.~\cite{Rosa:2017ury} and the outcome is the constant radio emissions. In other word, our investigation here is the precursor of the spherical evolution described in Ref.~\cite{Rosa:2017ury}. Moreover, it is shown in Fig.~\ref{uniform} that, for a set of benchmark parameters of astrophysical interest, the whole sphericalization process is of ${\cal O}$(1 hour), which is relatively very short on astronomical timescales. Therefore, if an axion cloud exist, it is highly probable to see it in a spherical configuration, no matter what its initial shape is. In other words, our present investigation of the non-spherical cloud evolution provides a warrant for adopting the spherical axion cloud for phenomenological studies in the literature~\cite{Rosa:2017ury}.

Here we would like to mention that the final spherical profile does not mean that the state is in the $s$-orbital quantum state. As argued in Ref.~\cite{Kofman:1997yn}, when the photons are abundantly produced, the quantum coherence in the original axion cloud would be lost soon due to the backreaction and rescattering effects from the created photons. As a result, the state cannot be viewed as a large coherent quantum state or a large classical field background any more. Rather, it is just a collection of non-coherent low-energy axion particles with the overall shape to be of spherical symmetry. The shape changing of the axion cloud is a consequence of the redistribution of the axion number density due to the combined effects of the stimulated emission and the superradiance. 

{In the above discussion, we simplify our discussion by ignoring several important effects, which might potentially change the evolution of the axion cloud and hamper the observed sphericalization. The first effect is the re-absorption of axions by the central black hole. One can estimate the absorption rate by assuming that only axion particles near the event horizon would be captured by the black hole. In the natural units with $\hbar = c = 1$, the axion number density can be taken to be around $n_a\sim 1/a_0^3 = (GM m_a)^3$ where the reduced Bohr radius of this system is $a_0 \equiv 1/(GMm_a)$. Hence, the axion flux density penetrating the black hole horizon is $\Gamma_{\rm ab} \approx \beta n_a (4\pi r_+^2)\approx 16\pi \beta (GMm_a)^5 m_a = 16\pi \beta \alpha^5 m_a$, where $r_+ \sim 2GM$ is of the size of the black hole horizon, $\beta \sim 0.1$ the typical axion velocity in the cloud, and $\alpha = GM m_a = M\mu$ the gravitational fine structure constant, respectively. By taking the typical values of the above quantities, the axion absorption rate is of order of $1 \times 10^{-5}$ Hz which is much smaller than the sphericalization rate of ${\cal O}(5\times 10^{-4})$ Hz as clearly shown in Eq.~\ref{uniform}. Thus, we expect that the absorption effect can be neglected for the evolution of the axion cloud as a whole.  

Another potentially important process is the axion self-interactions which might interfere with the resonance. In Ref.~\cite{Braaten:2019knj}, the cross section for the elastic scattering of two axions due to self-interactions is given by $\sigma_\text{SI} \approx 8\pi m_a/(32\pi f_a^2)$ for the conventional axion potential, while the stimulated emission cross section is calculated in \cite{Berera:2023fhd} with the result as $\sigma_\text{SE}\sim({\lambda_0/ 2\pi })^2   ( {  \lambda_0/\tau_a } )  $ , where $\lambda_0$ denotes the wavelength of the incident photon and $\tau_a$ as the axion lifetime. As a consequence, the ratio of these two cross sections is $\sigma_\text{SI}/\sigma_\text{SE}\sim10^{-47}$ with benchmark parameter values. In other words, comparing to the stimulated emissions, the axion self scatterings can be negligible, and could not generate any visible effects on the resonance.

	
Finally, the benchmark black hole mass is taken to be $M= 4\times 10^{-7} M_\odot$, which indicates that the black hole should be of primordial origin. Also, in order to generate the required high black hole spin for the axion superradiance, the formation of such a black hole requires multiple mergers. It has been demonstrated in Ref.~\cite{Scheel:2008rj} that such a black hole with its spin around 0.7 can be achieved by mergers of nonspinning binary black holes. Massive primordial black holes (PBH) are possibly grouped in dense clusters~\cite{Clesse:2015wea}. They can be captured in binaries and subsequently merge by losing their energies by radiating gravitational waves. The merging rate can reach~\cite{Clesse:2016vqa}
\begin{eqnarray}
	\tau_{\rm merg} \sim   7 \times 10^{-9} \,f_\text{DM} ~ \delta_\text{PBH}^\text{loc} \ \text{yr}^{-1} \text{Gpc}^{-3}\,,
\end{eqnarray}
where $f_\text{DM}$ and $\delta_\text{PBH}^\text{loc}$ are the PBH density fraction in dark matter and the local density enhancement due to the PBH clustering, respectively. If a large local PBH density contrast $\delta_{\rm PBH}^{\rm loc} \sim 10^9 - 10^{10}$ and a substantial PBH abundance with $f_{\rm DM} \sim 0.1$ are realized in nature, it is possible to expect a few to tens of PBH merger events to occur every year, which might lead to a high black hole spin with the mechanism in Ref.~\cite{Scheel:2008rj}. On the other hand, it was demonstrated in Ref.~\cite{Harada:2017fjm} that high-spin primordial black holes could even be directly formed in the matter-dominated era.

}

\section{Discussion}\label{SecDiscussion}

From the above discussion, the evolution of this axion cluster can be summarized as follows:
it begins with a low-density axion ``fog'' which becomes growing due to the superradiance from a center rotating black hole. The fastest growing mode corresponds to $(l,m)=(1,1)$ which indicates that there is an initial angular variation in the number density of axions. The stimulated emission efficiency strongly depends on the axion density: around the equatorial plane, the axion density is so high that the stimulated emission depletes the axion faster than the axion production rate of the superradiance, which decreases the axion density there. On the other hand, when approaching the polar region, less and less axions distributed there, so that the black hole superradiance dominates over the stimulated emission and the axion density becomes increased. {The combined effect of the stimulated emission and superradiance is that the axion cloud becomes spherically distributed, and the following evolution would be described by the spherical axion-photon system studied in Ref.~\cite{Rosa:2017ury}.  } 

\begin{figure}[ht]
        \centering
\includegraphics[width=0.5\textwidth]{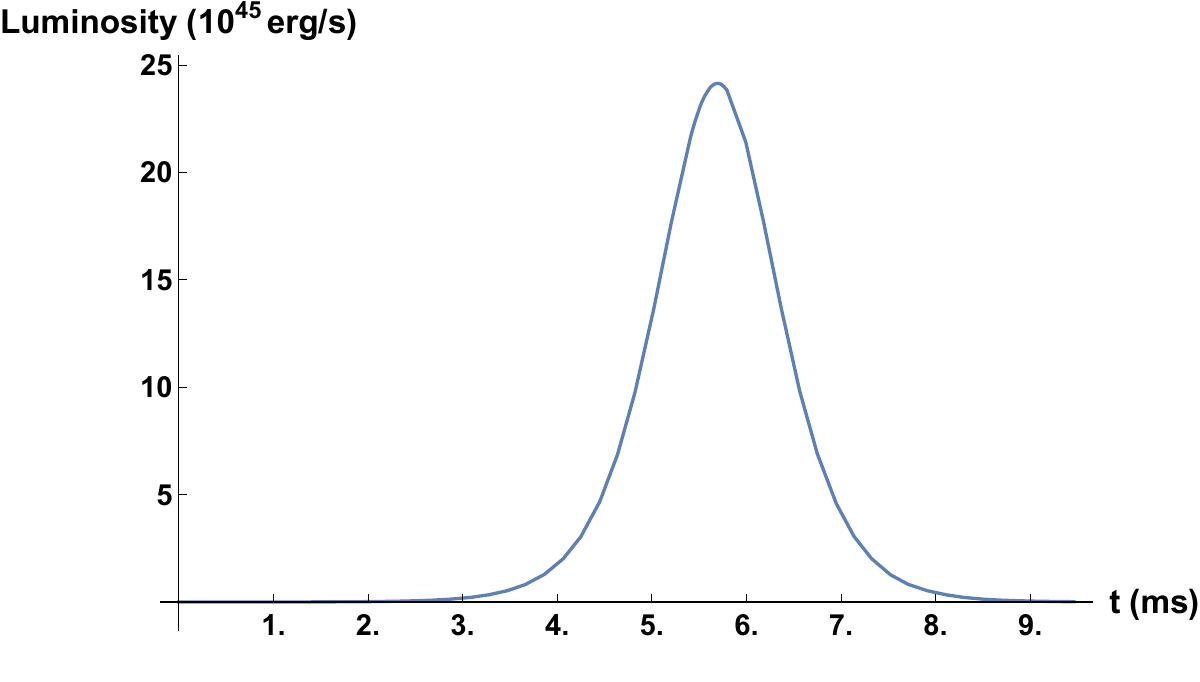}\includegraphics[width=0.5\textwidth]{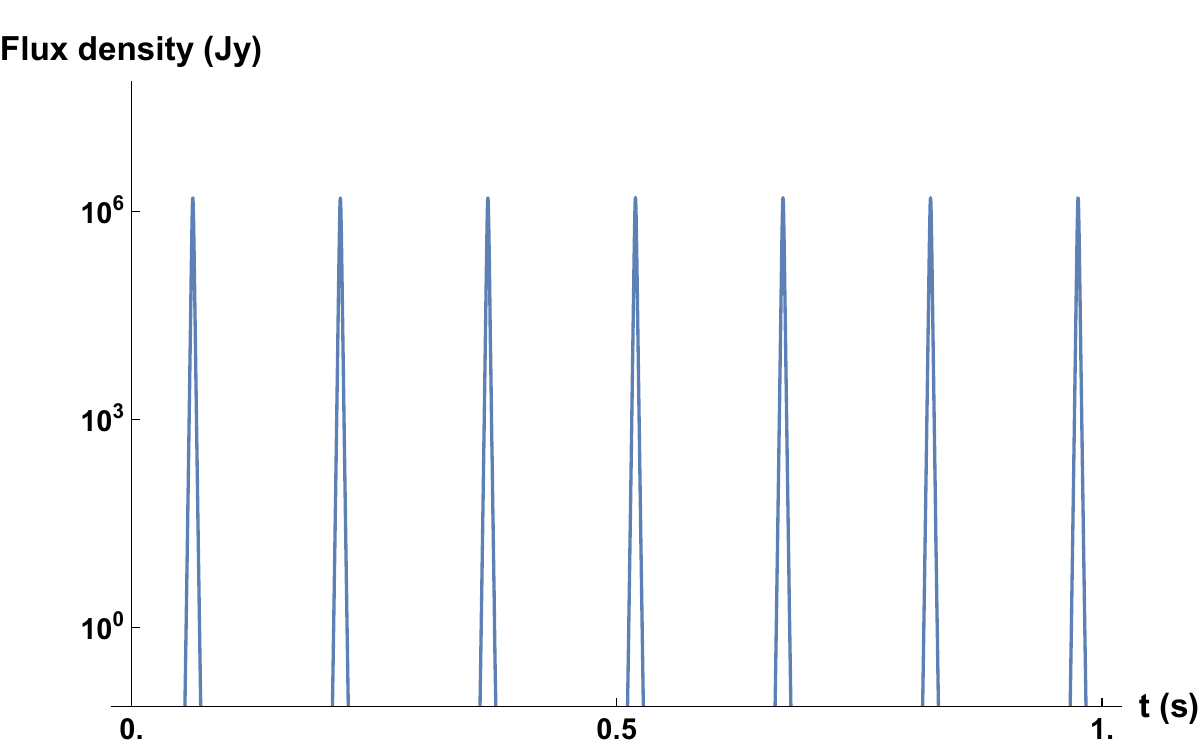}
\caption{\label{1pulse}
The luminosity of a typical pulse originated from one of the spikes on the left panel of Fig.~\ref{NSphSym} (left panel) and signals of the photon flux density received on the Earth for $m_a = 10^{-5}$~eV and $M=4\times10^{-7}M_\odot$ (right panel).}
\end{figure}
\begin{figure}[ht] \centering
\includegraphics[width=0.5\textwidth]{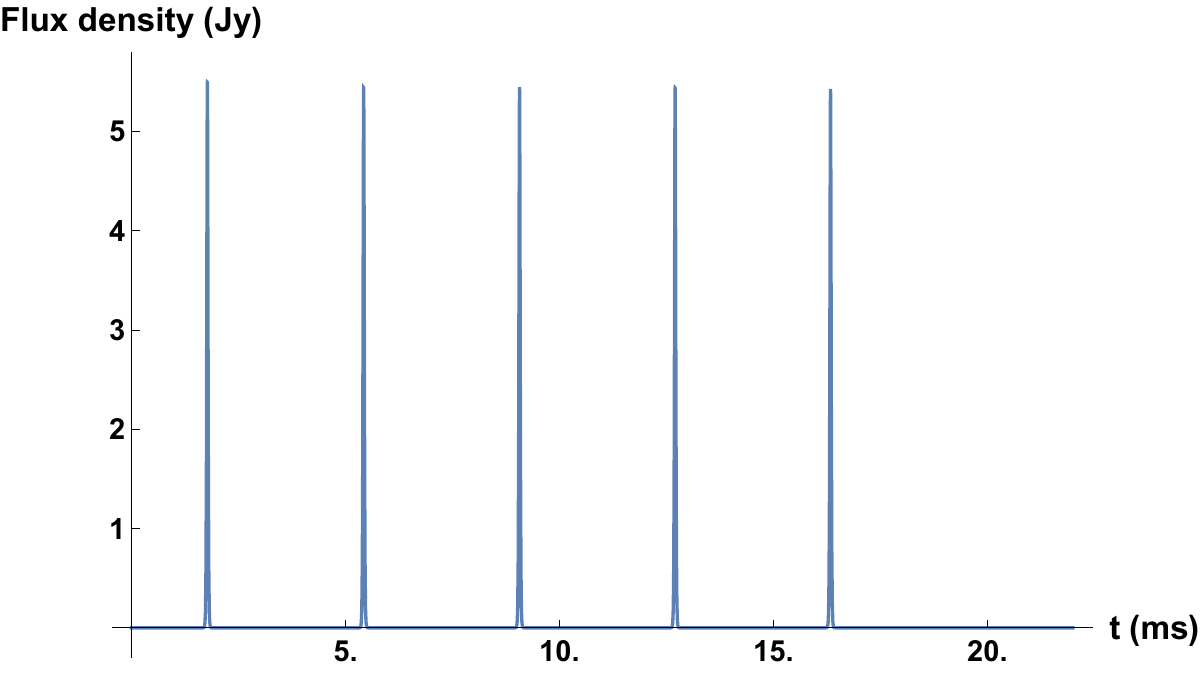}\includegraphics[width=0.5\textwidth]{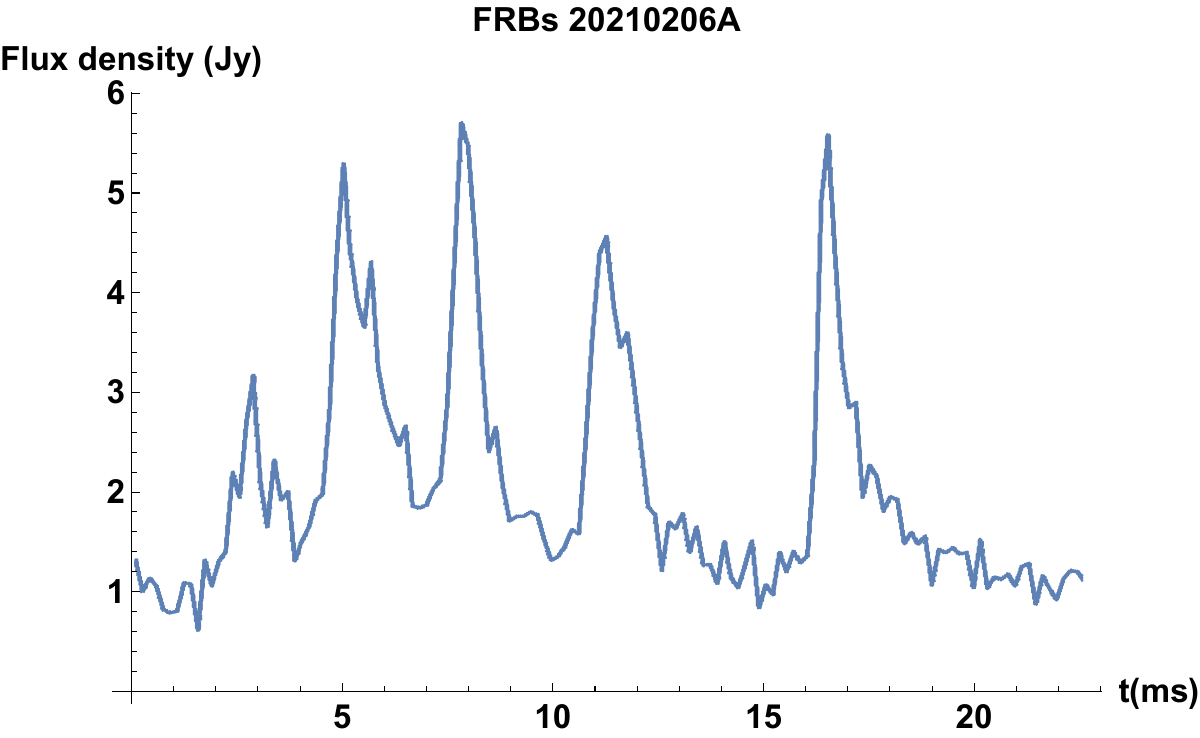}
\caption{\label{fluence}
The photon flux density received on the Earth for $m_a = 10^{-5}$~eV, $M = 7.25\times10^{-5}M_\odot$ and $K=10$ (left panel). In comparison, we show on the right panel the observed flux density of a repeating FRB 20210206A.}
\end{figure}

A wide variety of physical processes could produce transient astrophysical events on human-observable timescales. Fast radio bursts (FRBs) are one type of these transient events from unknown sources. {They are bright pulses with flux densities\footnote{Following the convention in the radio astronomy, we shall use the quantity called "flux density" to represent the magnitude of a FRB signal. The unit of this quantity is "Jansky" or "Jy" for short, which is equal to $10^{-26}$ watts per square meter per hertz (${\rm W/m^2/Hz}$). To get a feeling of the size of this unit, the radio flux density from the Sun is on the order of several thousand Janskys. } between 50 mJy and 100 Jy  at radio frequencies in the range of 0.4~GHz - 8~GHz. Their typical duration is of order milliseconds.  
In general, the FRBs can be classified into two categories: repeaters and non-repeaters. It has been argued that these two types of FRBs may have different origins~\cite{Palaniswamy:2017aze}. 
Our analysis suggests that pulses produced by axion clusters 
may provide one candidate for the repeater-like FRBs. }

We depict the typical spectrum for one single pulse during the uniformization process of the axion cloud on the left panel of  Fig.~\ref{1pulse}. It shows that the yielded pulse lasts about 2~ms with its frequency of 1200~MHz, and the latter quantity can be derived from half of the axion mass due to its origin from an axion decay. The total luminosity can be as high as $10^{45}$ erg/s, while total energy released for the pulse duration can reach beyond $10^{42}$ erg. Note that this energy is very close to a recent study of FRB 20220610A~\cite{Ryder:2022qpg} which showed the energy of a burst was $2\times10^{42}$ erg. 
We also show on the right panel of Fig.~\ref{1pulse} the signal strengths of a series of pulses in terms of the flux density. 
For a source placed in the luminosity distance of $\sim 6700$~Mpc corresponding to the cosmological redshift $z=1$, the fluence\footnote{The fluence is a measure of the energy received per unit area. In the radio astronomy, the proper unit is Jy-ms. } 
of one pulse is of order of $10^5 \sim 10^6$ Jy-ms, 
which indicates that this kind of periodical radio emissions should not be difficult to detect and identify.

Due to similarities in the duration and flux density, the pulses from the stimulated emission from axion clouds can be used to explain the observed FRBs. We can compare our predicted signals with the detailed information of FRBs given in the catalogs~\cite{CHIMEFRB:2021srp, Petroff:2016tcr}.  
For instance, at 1.4 GHz, a 30 Jy pulse of duration less than 5 ms was found and associated with an extragalactic singular event~\cite{Lorimer:2007qn}. Additional millisecond-duration FRBs centered at 1.38 GHz with redshifts of 0.5 to 1 were detected and the energy released ranges from $10^{31}$ J to $10^{33}$ J~\cite{Thornton:2013iua}, which are close to our estimated energy on the left panel of Fig.~\ref{1pulse}. Eight repeating bursts from FRB 121102 were detected in a single day~\cite{Spitler:2016dmz} in 2015. It was argued earlier that the possible sources can be low-luminosity active galactic nucleus~\cite{Chatterjee:2017dqg} or a young neutron star energizing a supernova remnant~\cite{Marcote:2017wan}. It is interesting to note that the temporal locations of pulses plotted on the right panel of Fig.~\ref{1pulse} is consistent with this event, implying a possible connection between the axion cloud emission and this FRB signal. Unfortunately, the estimate on the photon flux density seems to be much larger than this signal. However, it is possible to match the observed signal by adjusting model parameters. 
For example, if we set $m_a =10^{-5}$~eV and the black hole mass $M= 7.25\times10^{-5}\,{M_\odot }$ while letting $K=10$ in Eq.~\eqref{lifetime}, the yielded photon pulses are shown on the left plot of Fig.~\ref{fluence}, which has already been very similar in the amplitude and cadence as another observed repeating FRB event 20210206A~\cite{CHIMEFRB:2021fvq} on the right plot. In Table.~\ref{table1}, we provide some details of observational FRB data and our model predictions for comparison. 

{Still, there are two main differences between the left and right-hand plots in Fig.~\ref{fluence}. One is the apparent missing pulse in the left spectrum, since the time lag between the fourth and fifth pulses is about twice of previous intervals. This feature can be understood in terms of the precession of the axion cloud and its associated black hole. For instance, due to the relatively slow precession, the polar regions of the cloud might point to the direction of the Earth, causing the sudden decrease of the pulse amplitude and an undetectable signal. Another interpretation might involve an accidental event in which a fast-moving object travels perpendicular to the line of sight, blocks this pulse, and reflects it towards other directions. Furthermore, the other distinction manifests in the pulse broadness in our predictions and real observations. It is clear that the pulses are predicted to be of short duration, while the measured signals are much wider. In the literature, there are multiple effects to broaden the electromagnetic pulses. For example, it was pointed in Refs.~\cite{Cordes:2002wz,Petroff:2019tty,Thornton:2013iua} that the signals may scatter off free electrons and other charged particles when they pass through the interstellar medium, which might lead to the dispersion and the broadening of the signal. Or, as in Refs.~\cite{Rickett:1990dr}, one might think that the axion cloud is surrounded by the dense or turbulent environments, where electromagnetic waves might undergo interference or multipath propagation. This could also lead to a spread in arrival times. Fully explaining all the substructures of the observed spectrum would require us to consider many astrophysical situations and effects, which is obviously beyond the scope of the present work. In fact, the study of FRB substructures is active among the community~\cite{Petroff:2021wug}.}

\begin{table}[ht]
\begin{center}
\begin{tabular}{  c|c|c|c|c } 
    &     & duration & fluence/flux density & frequency \\\hline
theory & $m_a = 10^{-5}$~eV, $M\sim4\times10^{-7}M_\odot$  & 2 ms &  $10^6$ Jy  & 1.2 GHz \\ 
    & $m_a = 10^{-5}$~eV, $M\sim7.25\times10^{-5} M_\odot$   & 0.1 ms & 5.4 Jy & 1.2 GHz \\\hline
observations    &FRB 010824\,\cite{Lorimer:2007qn} & 5 ms & 30 Jy & 1.4 GHz\\
    &FRB 110220\,\cite{Thornton:2013iua}  & 0.8 ms & 8 Jy-ms & 1.3 GHz \\
    &FRB 121102\,\cite{Spitler:2016dmz}(repeater)   & 3 to 8 ms & 0.06 to 1 Jy-ms  & 1.4 GHz\\
    &FRB 20210206A\,\cite{CHIMEFRB:2021fvq}(repeater)   & 1 to 2 ms & 5 to 6 Jy  & 0.4 to 0.8 GHz
\end{tabular}
\caption{\label{table1}
Comparison of characteristics between the pulses predicted by the stimulated emission in axion clouds and those from a few FRB measurements. In the 4th column, the flux density of FRB 010824 and FRB 20210206A and the fluences of other two FRB events are given. }
\end{center}
\end{table}

{We close this section by mentioning several other works trying to relate the observations of FRB signals to the axion models. Refs.~\cite{Iwazaki:2014wka, Iwazaki:2018squ} proposed that the collision of a neutron star and an axion star would be the origin of FRBs. The idea of generating FRBs through an axion star moving in the magnetosphere of a neutron star has been investigated in Ref.~\cite{Buckley:2020fmh}. 
Some further FRB generation mechanisms involving axions have been studied in Refs.~\cite{Tkachev:2014dpa,vanWaerbeke:2018nyj}. Moreover, FRB measurements can also provide constraints on the parameter space for axions or axion like particles, as shown in Ref.~\cite{Caputo:2019tms}. Finally, Ref.~\cite{Prabhu:2023cgb} proposed the opposite situation that FRBs may convert to axion bursts.}


\section{Conclusion}\label{SecConc}
We have presented an updated evolution model of the stimulated emission from a non-spherical axion cluster, specifically for the initial distribution corresponding to the axion growth mode $(l,m)=(1,1)$ from the black hole superradiance. In the case of a sourceless axion cluster, 
the stimulated photon emission is the strongest on the equatorial plane and becomes weaker towards the polar regions. As a result, the entire process releases only one pulse of photons. On the other hand, when the black hole superradiance acts as an axion production source, the simulated emission and the superradiant growth would interplay with each other, and finally reach a balance point where the axion cloud becomes spherically distributed. 
Numerous pulses are released during this uniformization process and the axion-photon system enters a quasi-stable stage with constant radio emissions, which has already well-studied in the literature. {Thus, 
since the sphericalization process is relatively short on astrophysical timescales,
our present investigation may provide a warrant for adopting the spherical configuration as a good approximation to the actual axion profile in most cases.}
The energy and temporal characteristics of the yielded photon pulses could be applicable to the observed phenomena of FRBs.

\begin{acknowledgments}
LC gratefully thanks Thomas W. Kephart at Vanderbilt University for the discussions on this work.
This work is supported in part by the National Key Research and Development Program of China (Grant No. 2020YFC2201501 and 2021YFC2203003), and the National Natural Science Foundation of China (NSFC) under Grant No. 12347103.
\end{acknowledgments}

\end{document}